\documentclass[aps,reprint,groupedaddress]{revtex4-1}
\usepackage{mathrsfs}
\usepackage{graphicx}
\usepackage{dcolumn}
\usepackage{bm}
\usepackage[utf8]{inputenc}
\usepackage[T1]{fontenc}
\usepackage{booktabs, array, float, tabularx, booktabs, lipsum, amsmath, multirow}
\usepackage{siunitx, xcolor}
\usepackage[version=4]{mhchem}
\usepackage[breaklinks,colorlinks,linkcolor=blue,citecolor=blue,urlcolor=blue]{hyperref}

\newcommand{\be}{\begin{equation}}
\newcommand{\ee}{\end{equation}}
\newcommand{\bea}{\begin{eqnarray}}
\newcommand{\eea}{\end{eqnarray}}

\begin{document}

\title{Chiral coupling between a ferromagnetic magnon and a superconducting qubit}
\author{Ya-long Ren}
\affiliation{MOE Key Laboratory for Nonequilibrium Synthesis and Modulation of Condensed Matter,
Shaanxi Province Key Laboratory of Quantum Information and Quantum Optoelectronic Devices, School of physics,
Xi'an Jiaotong University, Xi'an 710049, China}

\author{Sheng-li Ma}
\email{msl1987@xjtu.edu.cn}
\affiliation{MOE Key Laboratory for Nonequilibrium Synthesis and Modulation of Condensed Matter, Shaanxi Province Key Laboratory of Quantum Information and Quantum Optoelectronic Devices, School of physics,
Xi'an Jiaotong University, Xi'an 710049, China}

\author{Fu-li Li}
\email{flli@mail.xjtu.edu.cn}
\affiliation{MOE Key Laboratory for Nonequilibrium Synthesis and Modulation of Condensed Matter,
Shaanxi Province Key Laboratory of Quantum Information and Quantum Optoelectronic Devices, School of physics,
Xi'an Jiaotong University, Xi'an 710049, China}

\begin{abstract}
Chiral coupling at the single-quantum level promises to be a remarkable potential for quantum information processing. Here we propose to achieve a chiral interaction between a magnon mode in a ferromagnetic sphere and a superconducting qubit mediated by a one-dimensional coupled-cavity array. When the qubit is coupled to two lattice sites of the array and each one is encoded with a tunable phase, we can acquire a directional qubit-magnon interaction via the quantum interference effect. This work opens up a new route to construct chiral devices, which are expected to become a building block in quantum magnonic networks.
\end{abstract}

\maketitle

\section{Introduction}

In recent years, chiral phenomena have been attracting intense attentions in quantum optics \cite{FA1,Xi2021,Owens2022}.
Chirality refers to the breaking of inversion symmetry \cite{F1,YuT2022},
and plays very important roles in various fields of science and technology,
such as molecule detection \cite{NA2019,DR2022,Das2022}, optical communication
and information processing \cite{Ramos2016,Coles2016}.
A central issue in this subject is to tailor chiral couplings \cite{F4,F5,Soellner2015,F6},
which can be used to control the directionality of the spontaneous emission and modify the interactions
between multiple quantum emitters \cite{F2}. These capabilities create the
possibilities of one-way information flow, deterministic quantum state
transfer, and quantum simulation of many-body physics \cite{Roushan2016,F10,F11}.
To generate such a chiral coupling, a variety of physical processes have been explored,
such as optomechanical interactions \cite{G1,G2}, Brillouin scattering \cite{G3,G4},
synthetic magnetic field \cite{G5,G6}, and topological engineering \cite{G7,G8}.

On the other hand, hybrid quantum systems involving the integration of
ferromagnetic materials and superconducting circuits have achieved rapid
progress very recently \cite{Jing2021,Shen2021,Kani2022}. First, they are
compatible with each other to achieve coherent light-matter interaction \cite{Rameshti2022}.
Moreover, ferromagnetic materials, such as yttrium iron garnet (YIG),
have spin density many orders of magnitude higher than dilute spin ensembles
\cite{Huebl2013}. As a result, strong and ultrastrong coupling of
ferromagnetic magnons to microwave photons in a superconducting cavity has
been experimentally demonstrated \cite{C1,C2,Zhang2015,Hou2019}. In such hybrids, many
intriguing phenomena have been studied, including magnon nonclassical states
\cite{D1,D2,D6,D7,D8,D9,D10,Guan2022}, Floquet engineering \cite{D11}, spin currents \cite{D12,D13},
magnon-induced nonreciprocity \cite{D14,D15,D16,D17,D18,RenR},
non-Hermitian physics \cite{Zhang2017,D20,Liu2019,D21,D22,D24}, and dark matter detection \cite{D25,D26,D27}.

\begin{figure*}[tbph]
\centerline{\includegraphics[width=15cm]{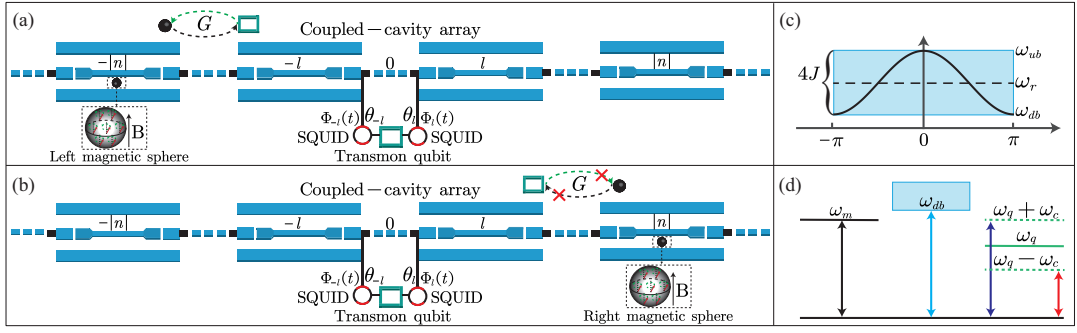}}
\caption{Schematic diagram of the hybrid system.
A superconducting transmon qubit is coupled to two lattice sites of a coupled-cavity array
via superconducting quantum interference devices (SQUIDs).
A magnetic sphere supporting the magnon modes is placed to the (a) left-side cavity $-|n|$ or (b) right-side cavity $|n|$,
and the resonance frequency of magnon mode can be tuned by adjusting the external magnetic field $B$.
(c) Propagating frequency band of the coupled-cavity array.
(d) Two side-band processes of the superconducting qubit are induced by the time-dependent modulation coupling.}
\end{figure*}
Apart from the photon-magnon polariton, coherent coupling between a magnon
mode and a superconducting qubit can also be mediated via a superconducting
cavity \cite{E2}. Such a coupled qubit-magnon system is particularly
appealing, because the superconducting qubit with a strong anharmonicity can
be used to explore spintronics and magnonics in the quantum limit \cite{B4}. And many
theoretical schemes for the preparation of single magnon sources \cite{E8,E9,E10},
magnon-magnon entangled states \cite{E11,E12,E14}
and magnonic cat states \cite{Sharma20222,Kounalakis20222} have been proposed based on this composite
system. In a recent experiment \cite{E6}, Quirion et al. reported the
detection of a single magnon in a millimeter-sized YIG sphere with a quantum
efficiency of up to 0.71. Although exciting progresses have been made in
this area, to the best of our knowledge, the possibility of a chiral
qubit-magnon coupling has not been revealed.

In this paper, we propose a novel mechanism for the realization of a chiral coupling
between a magnon mode hosted by a ferromagnetic sphere and a superconducting qubit
through a one-dimensional coupled-cavity array.
Here, the qubit is designed to interact with the cavity array via two different sites,
and each one is encoded with a tunable phase.
After adiabatically eliminating the degree of freedom of the cavity array,
we find that the effective qubit-magnon coupling becomes phase-dependent.
In particular, by manipulating those nontrivial phases, we can acquire a chiral qubit-magnon interaction, which stems from the
quantum interference effect. To be specific, the qubit-magnon coupling is formed in
a given direction whereas it is forbidden in the opposite direction. Based
on this chirality, many intriguing quantum applications can
be implemented, such as the chiral qubit-magnon entanglement and the
directional magnon blockade. In a broader view, the proposed chiral
qubit-magnon coupling opens a new perspective for the design and exploration
of chiral magnonic devices, which could promote a variety of practical applications in the field of quantum magnonics.

\section{Theoretical model}

As sketched in Fig. 1, we consider a hybrid quantum system that consists of
a one-dimensional coupled-cavity array, a superconducting qubit,
and a single-crystalline YIG sphere. The cavity array is made up of $N$
linearly coupled superconducting transmission line cavities,
and we focus here on the thermodynamic limit $N\gg1$ for simplicity \cite{Zhou2008,Fang2012}.
The superconducting cavity can be modeled as a simple harmonic oscillator with the effective
capacitance $L_{r}$ and inductance $C_{r}$. Then, the Hamiltonian of the
coupled-cavity array yields (hereafter we set $\hbar=1$)
\begin{equation}
H_{r}=\omega_{r}\sum_{z}a^{\dag}_{z}a_{z}+J\sum_{z}(a^{\dag}_{z}a_{z+1}+a_{z}a^{\dag}_{z+1}),
\end{equation}
where $\omega_{r}=1/\sqrt{L_{r}C_{r}}$ is the resonance
frequency of superconducting cavity, and $a_{z}$ ($a^{\dag}_{z}$) is the photon annihilation (creation)
operator on lattice site $z$. Besides, $J$ is the photon hopping rate
between two neighboring cavities. Experimentally, the strong coupling
between two superconducting cavities has been achieved by connecting them
with a coupler, such as a capacitor or a SQUID loop of Josephson-junction
circuit \cite{H1,H2,H3}.

A magnetic sphere is coupled to the $n$th microwave cavity via the magnetic-dipole interaction.
For convenience, we mark the magnetic sphere as left (right) magnetic sphere when $n\leq-1$ ($n\geq1$).
The magnetic sphere supports a series of magnetostatic modes,
and we are only interested in the fundamental magnon mode,
i.e., a uniform collective mode that all the spins precess in phase.
Hence, the associated Hamiltonian reads
\begin{equation}
\begin{aligned}
H_{m}=\omega_{m}m^{\dag}m+g_{m}(a_{n}^{\dag}m+a_{n}m^{\dag}).
\end{aligned}
\end{equation}
In the above equation, $\omega_{m}=\gamma_{e}B$ is the resonance frequency of the magnon mode,
where $\gamma_{e}/2\pi=28$ GHz/T is the electron gyromagnetic ratio,
and $B$ is the external magnetic field. $m$ ($m^{\dag}$) is the magnon annihilation (creation) operator.
In addition, $g_{m}=\eta\gamma_{e}\sqrt{N_{s}\mu_{0}\hbar\omega_{r}/(4V_{r})}$ denotes the magnon-photon coupling strength \cite{C3},
where $N_{s}$ is the number of the net electron spins in the magnetic sphere, $\mu_{0}$ is the vacuum permeability,
and $V_{r}$ is the mode volume of the microwave cavity.
The coefficient $\eta$ is determined by the position of the magnetic sphere inside the microwave cavity.

Finally, we consider that a superconducting qubit is coupled to the cavity array via two different sites \cite{Wangxin,Wang2021,Du20222}.
In practical experiment, we can adopt a transmon-type qubit as a concrete example \cite{Koch2007,Andersson2019,Kannan2020},
which interacts with the $-l$th and $l$th cavities via two SQUIDs, simultaneously.
As a result, the interaction Hamiltonian takes the form
\begin{equation}
\begin{aligned}
H_{q}=\omega_{q}\sigma_{+}\sigma_{-}+\sum_{j=-l,l}\tilde{g}_{q_{j}}(t)(a_{j}^{\dag}\sigma_{-}+a_{j}\sigma_{+}),
\end{aligned}
\end{equation}
where $\omega_{q}$ is the transition frequency of the superconducting qubit between
the excited state $|e\rangle$ and the ground state $|g\rangle$, and
$\sigma_{-}=|g\rangle\langle {e}|$ ($\sigma_{+}=(\sigma_{-})^{\dag}=|e\rangle\langle {g}|$)
is the usual Pauli operator. Experimentally, by periodically modulating the external flux
$\Phi_{j}(t)$ threading the SQUID loop \cite{Felicetti2014,Lu2017},
one can get the time-dependent coupling strength
$\tilde{g}_{q_{j}}(t)=2g_{q_{j}}\mathrm{cos}(\omega _{c}t-\theta _{j})$,
in which $g_{q_{j}}$ is the amplitude, $\omega_{c}$ is the modulation frequency,
and $\theta_{j}$ is the modulation phase.

At present, the total Hamiltonian of the whole system is given by
\begin{equation}
H=H_{r}+H_{m}+H_{q}.
\end{equation}
To go a further step, we perform the Fourier transformation $a_{k}=\sum_{z}a_{z}e^{ikz}/\sqrt{N}$
($k\in[-\pi,\pi]$), and $H$ can be rewritten in the momentum representation as
\begin{widetext}
\begin{equation}
\mathcal{H}=\sum_{k}\omega_{k}a^{\dag}_{k}a_{k}+\omega_{m}m^{\dag}m+\omega_{q}\sigma_{+}\sigma_{-}
+\sum_{k}\Big[\frac{g_{m}e^{ikn}}{\sqrt{N}}a_{k}^{\dag}m
+\sum_{j=-l,l}\frac{2g_{q_{j}}\mathrm{cos}(\omega_{c}t-\theta_{j})e^{ikj}}{\sqrt{N}}a_{k}^{\dag}\sigma_{-}+\mathrm{H.c.}\Big],
\end{equation}
\end{widetext}
where $\omega_{k}=\omega_{r}+2J\mathrm{cos}(k)$ is the dispersion relation of the coupled-cavity array,
which is centered at $\omega_{r}$ with the up band edge $\omega_{ub}=\omega_{r}+2J$
and the down band edge $\omega_{db}=\omega_{r}-2J$  [see Fig. 1(c)].
In order to obtain a desired interaction, we consider that the qubit's frequency is far-detuned from the down band edge [see Fig. 1(d)].
At the same time, the modulation frequency is chosen to satisfy $\omega_{q}+\omega_{c}=\omega_{m}\gg\{g_{m},g_{q_{j}}\}$,
such that the qubit is effectively coupled to the cavity array through the blue-side band process.
By performing the rotating wave approximation to neglect the fast oscillating terms (see Appendix A for details),
the total Hamiltonian is simplified as
\begin{widetext}
\begin{equation}
\mathcal{H}=\sum_{k}\omega_{k}a^{\dag}_{k}a_{k}+\omega_{m}m^{\dag}m+\omega_{m}\sigma_{+}\sigma_{-}
+\sum_{k}\Big[\frac{g_{m}e^{ikn}}{\sqrt{N}}a_{k}^{\dag}m
+\sum_{j=-l,l}\frac{g_{q_{j}}e^{i(\theta_{j}+kj)}}{\sqrt{N}}a_{k}^{\dag}\sigma_{-}+\mathrm{H.c.}\Big],
\end{equation}
\end{widetext}
where two tunable phases are encoded in the coupling strengths between the qubit and the cavity array.
In our scheme, the cavity array as a data bus is utilized to indirectly couple the qubit and the magnon mode,
so that the qubit-magnon coupling strength will naturally inherit those phase information.
By properly choosing those phases, we can achieve a chiral qubit-magnon interaction.

\section{Chiral qubit-magnon coupling}

In this section, we show how to realize the chiral qubit-magnon coupling originating from the quantum interference effect.
In the dispersive regime $\Delta=\omega_{db}-\omega_{m}\gg{\{g_{m},g_{q_{j}}\}/\sqrt{N}}$,
the coherent qubit-magnon interaction can be induced by the exchange of virtual photons.
After adiabatically eliminating the degree of freedom of the cavity array (see Appendix B for details),
we can obtain the effective Hamiltonian as
\begin{equation}
\mathcal{H}_{e}=\omega_{m}^{\prime}m^{\dag}m+\omega_{q}^{\prime}\sigma_{+}\sigma_{-}
-Gm^{\dag}\sigma_{-}-G^{*}m\sigma_{+},
\end{equation}
where
\begin{subequations}
\begin{align}
\omega_{m}^{\prime}&=\omega_{m}-\frac{g_{m}^{2}}{\Delta\sqrt{\mathrm{coth}(1/2\lambda)}}, \\
\omega_{q}^{\prime}&=\omega_{m}-\frac{g_{q_{-l}}^{2}+g_{q_{l}}^{2}
+2\mathrm{cos}(\theta_{-l}-\theta_{l})g_{q_{-l}}g_{q_{l}}e^{\frac{-|2l|}{\lambda}}}
{\Delta\sqrt{\mathrm{coth}(1/2\lambda)}}, \\
G&=\sum_{j=-l,l}G_{j}=\sum_{j=-l,l}\frac{(-1)^{|n-j|}g_{q_{j}}g_{m}e^{i\theta_{j}}e^{\frac{-|n-j|}{\lambda}}}
{\Delta\sqrt{\mathrm{coth}(1/2\lambda)}}.
\end{align}
\end{subequations}
Here, $\omega_{m}^{\prime}$ ($\omega_{q}^{\prime}$) is the effective frequency of the magnon mode (qubit).
$\lambda=\frac{1}{\mathrm{arccosh}(1+\Delta/2J)}$ is the correlation length
that characterizes the interaction range between the magnon mode and the qubit.
$G$ is the effective qubit-magnon coupling strength mediated by the coupled-cavity array.
It can be seen that since the qubit is simultaneously coupled to two lattice sites of the cavity array,
the qubit-magnon coupling $G$ contains two components of $G_{-l}$ and $G_{l}$, each of which carries a nontrivial phase.
It is because of these two nontrivial phases $\theta_{-l}$ and $\theta_{l}$, the perfect chirality can be created via quantum interference.

In order to better reveal the chirality and simplify our discussion, we only concentrate on the simplest case $l=1$,
i.e., the qubit is coupled to the cavity array via the -1th and 1th sites. Then, the qubit-magnon coupling $G$ becomes
\begin{equation}
G=\frac{(-1)^{|n-1|}g_{m}}{\Delta\sqrt{\mathrm{coth}(1/2\lambda)}}
\sum_{j=-1,1}g_{q_{j}}e^{i\theta_{j}}e^{\frac{-|n-j|}{\lambda}}.
\end{equation}
Obviously, the modulation phases $\theta_{-1}$ and $\theta_{1}$ have a significant influence on $G$,
and play a key role in generating the chiral coupling.
To clarify the physical mechanism, we first investigate the situation that the magnetic sphere is located on the right-side cavity $n\geq1$,
and the qubit-magnon coupling becomes $G\propto(g_{q_{-1}}e^{\frac{-2}{\lambda}}e^{i\theta_{-1}}
+g_{q_{1}}e^{i\theta_{1}})e^{\frac{-(n-1)}{\lambda}}$.
If we further set $g_{q_{1}}=g_{q_{-1}}e^{\frac{-2}{\lambda}}$
and let the modulation phases meet $\theta_{1}-\theta_{-1}=\pi$,
the qubit-magnon coupling will disappear with $G=0$.
This results from the destructive quantum interference.
In stark contrast,
\begin{figure}[tbph]
\centerline{\includegraphics[width=8cm]{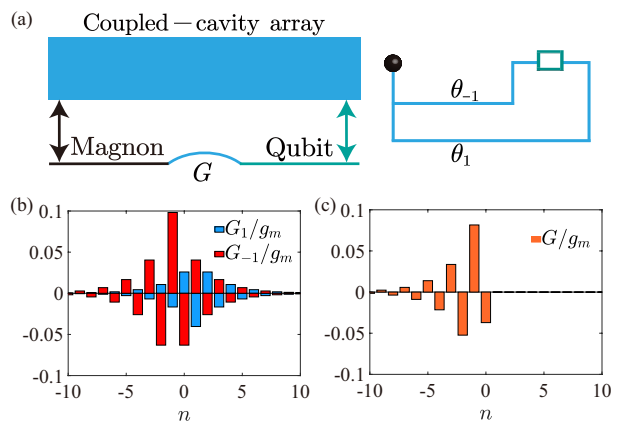}}
\caption{(a) Chiral qubit-magnon coupling mediated by a coupled-cavity array.
(b,c) Qubit-magnon coupling strength versus the position $n$ of the magnetic sphere,
where we set $\Delta/2\pi=100$ MHz, $J/2\pi=500$ MHz, $g_{m}/2\pi=41$ MHz, $g_{q_{-1}}/2\pi=45$ MHz,
$g_{q_{1}}/2\pi=18.5$ MHz, $\theta_{-1}=0$ and $\theta_{1}=\pi$.}
\end{figure}
\begin{figure*}[tbph]
\centerline{\includegraphics[width=15cm]{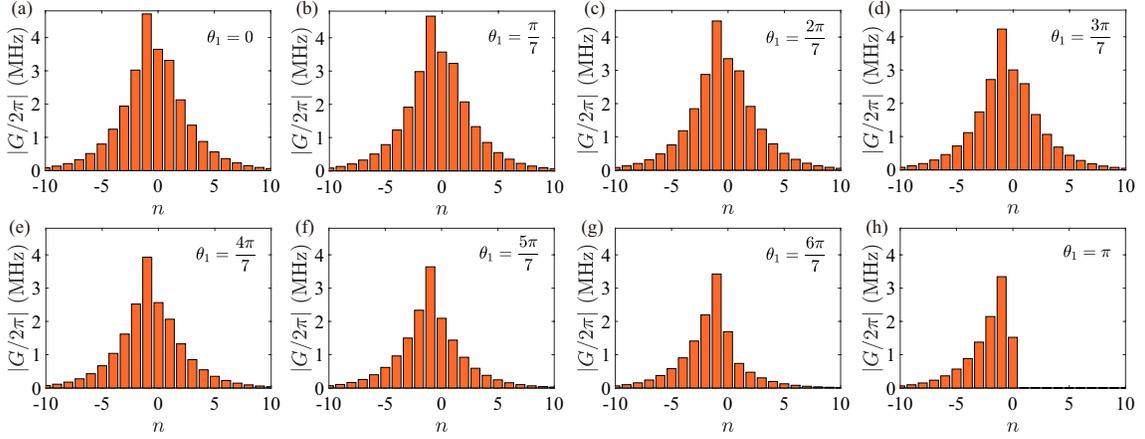}}
\caption{Qubit-magnon coupling strength $|G/2\pi|$ versus the position $n$ of the magnetic sphere for the different phases $\theta_{1}$,
where we take $\theta_{-1}=0$.
Other parameters are the same as in Fig. 2.}
\end{figure*}
when the magnetic sphere is located on the left-side cavity $n\leq-1$,
the two amplitudes $G_{-1}$ and $G_{1}$ can not cancel each other; that is, a nonzero coupling can be obtained with $G\propto(g_{q_{-1}}e^{\frac{2}{\lambda}}e^{i\theta_{-1}}
+g_{q_{1}}e^{i\theta_{1}})e^{\frac{n-1}{\lambda}}\neq0$.
Therefore, we can acquire a chiral qubit-magnon coupling through the quantum interference effect.
Remarkably, this chiral interaction can be controlled on-demand in terms of the potential operability and tunability of superconducting circuits.

We recall here that when the qubit interacts with the cavity array via a single site with $l=0$, the mediated qubit-magnon coupling yields
\begin{equation}
G=\frac{(-1)^{|n|}2g_{q_{0}}g_{m}e^{i\theta_{0}}e^{\frac{-|n|}{\lambda}}}{\Delta\sqrt{\mathrm{coth}(1/2\lambda)}},
\end{equation}
which has been discussed in Ref. \cite{E10}. Obviously, there is no chiral coupling for this case, i.e.,
the coupling of the qubit to the left magnetic sphere is the same as the coupling of the qubit to the right one.
In this work, the two-site couplings and the encoded nontrivial phases are exploited to create the desired chirality;
that is, the qubit-magnon coupling is formed in a chosen direction but vanished in the other.
Therefore, compared with the previous work \cite{E10}, the present one makes a significant step forward.

To clearly show the chiral coupling, we perform numerical simulations by considering some concrete parameters \cite{Leek2010,Lu2017,PhysRevX.12.031036}:
$\omega_{r}/2\pi=10.1$ GHz, $\omega_{m}/2\pi=9$ GHz, $\omega_{q}/2\pi=7$ GHz, and $\omega_{c}/2\pi=2$ GHz.
Additionally, we adopt the coupling strengths $g_{m}/2\pi=41$ MHz, $g_{q_{-1}}/2\pi=45$ MHz, $g_{q_{1}}/2\pi=18.5$ MHz, and $J/2\pi=500$ MHz.
In Fig. 2, we plot the qubit-magnon coupling strength as a function of the position $n$ of the magnetic sphere,
where the modulation phases are set $\theta_{-1}=0$ and $\theta_{1}=\pi$.
As shown in Fig. 2(b), $G_{-1}$ ($G_{1}$) decays exponentially around $n=-1$ ($n=1$),
so the qubit-magnon coupling strength can be adjusted by placing the magnetic sphere at different cavity $n$.
Notably, due to a $\pi$ phase different between $\theta_{-1}$ and $\theta_{1}$, $G_{-1}$ and $G_{1}$ always have a opposite sign for any $n$.
As a result, a directional qubit-magnon interaction is induced by the destructive quantum interference.
As expected, it can be seen from Fig. 2(c) that
the qubit-magnon coupling strength is nonzero for $n\leq-1$, while it is almost zero for $n\geq1$.

To gain an insight into the quantum interference,
we proceed to study the effects of modulation phases on the qubit-magnon coupling strength.
For a fixed phase $\theta_{-1}=0$, we plot $|G/2\pi|$ versus the modulation phase $\theta_{1}$ in Fig. 3.
When the phase $\theta_{1}$ is tuned from 0 to $\pi$, we can observe that
the coupling strength between the qubit and the right magnetic sphere is more and more significantly suppressed.
Particularly in $\theta_{1}=\pi$, one can obtain the chiral qubit-magnon coupling with a high contrast,
i.e., $|G/2\pi|\approx\{3.345,0.003\}$ MHz for $n=\{-1,1\}$.

To quantitatively describe the chirality, we further introduce the chiral factor as
\begin{equation}
\chi=\frac{|G(-|n|)|-|G(|n|)|}{|G(-|n|)|+|G(|n|)|},
\end{equation}
where $G(-|n|)$ ($G(|n|)$) is the coupling strength between the qubit and the left (right) magnetic sphere.
Here, the chiral factor $\chi>0$ denotes a chiral qubit-magnon coupling,
and the limit $\chi\rightarrow1$ indicates a perfect chirality.
As shown in Fig. 4,
\begin{figure}[tbph]
\centerline{\includegraphics[width=8cm]{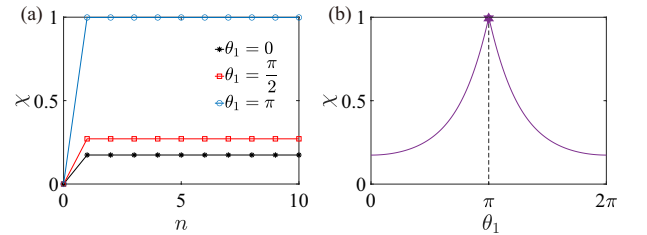}}
\caption{(a) Chiral factor $\chi$ versus the position $n$ of the magnetic sphere under the different phases $\theta_{1}$.
(b) Chiral factor $\chi$ as a function of the phase $\theta_{1}$ for $n=1$.
Other parameters are the same as in Fig. 2.}
\end{figure}
the chiral factor $\chi$ as a function of $n$ and $\theta_{1}$ is displayed.
Apart from $n=0$, the chiral factor can reach a maximum value,
which is independent of the position $n$ of the magnetic sphere [see Fig. 4(a)].
Additionally, by tuning the phase $\theta_{1}$ from 0 to $\pi$,
the maximum value of the chiral factor $\chi$ can approach $1$, as depicted in Fig. 4(b).
Therefore, our scheme exhibits a perfect chiral feature.

\section{Chiral phenomena}

In this section, we turn our attention to some chiral phenomena by exploiting the directional qubit-magnon interaction,
which may have many potential applications in quantum information processing.
To facilitate the following discussion, we take $\omega_{q}^{\prime}=\omega_{m}^{\prime}$ and $n=\pm1$ of the Hamiltonian $\mathcal{H}_{e}$ in Eq. (7).

\subsection{Chiral entanglement dynamics}

First, we demonstrate the chiral qubit-magnon entanglement based on the directional interaction.
Without loss of generality, the coupled qubit-magnon system is initially prepared in the separate state $|\Psi(t=0)\rangle=|0,e\rangle$,
i.e., the magnon mode is in the ground state $|0\rangle$ and the qubit is populated to the excited state $|e\rangle$.
In the absence of the dissipative process,
we can figure out $|\Psi(t)\rangle=\mathrm{cos}(Gt)|0,e\rangle+i\mathrm{sin}(Gt)|1,g\rangle$ exactly by solving the Schrödinger equation
$\mathcal{H}_{e}|\Psi(t)\rangle=i|\dot{\Psi}(t)\rangle$.
At the time $t=\pi/(4G)$, we can obtain a maximally entangled state
$|\Psi(t)\rangle=(|0,e\rangle+i|1,g\rangle)/\sqrt{2}$,
which is known as the Einstein-Podolsky-Rosen state \cite{Walgate2000}.

For the practical situation, the energy loss is inevitable in experiments
and the dissipation has to be taken into account.
For an open quantum system with a Markovian environment,
the dynamics of the hybrid qubit-magnon system is described by the quantum master equation
\begin{equation}
\begin{aligned}
\dot{\rho}=&-i[\mathcal{H}_{e},\rho]+\gamma_{q}\mathscr{L}[\sigma_{-}]\rho+\gamma_{m}\mathscr{L}[m]\rho.
\end{aligned}
\end{equation}
where we have assumed the zero working temperature, and neglected the thermal excitations. In Eq. (12), $\rho$ is the density matrix, and $\mathscr{L}[o]\rho=o\rho{o^{\dag}}-(o^{\dag}o\rho+\rho{o^{\dag}o})/2$
($o=\sigma_{-}$, $m$) is the standard Lindblad operator for a given operator $o$.
Besides, $\gamma_{q}$ ($\gamma_{m}$) represents the energy damping rate of the qubit (magnon mode).

\begin{figure}[tbph]
\centerline{\includegraphics[width=8cm]{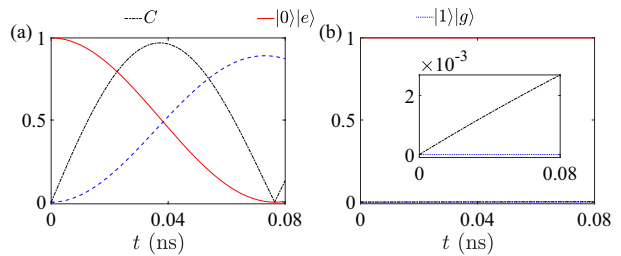}}
\caption{Time evolution of the concurrence and populations for the qubit and magnon mode,
where we take $\gamma_{q}/2\pi=0.05$ MHz and $\gamma_{m}/2\pi=0.5$ MHz.
[(a) $n=-1$; (b) $n=1$]. Other parameters are the same as in Fig. 2.}
\end{figure}
In the basis of \{$|1,e\rangle$, $|0,e\rangle$,$|1,g\rangle$, $|0,g\rangle$\},
the formal solution of the density operator $\rho$ for the qubit and the magnon mode takes the form
\begin{equation}
  \rho= \left(
 \begin{matrix}
   0&               0&                  0&                             0  \\
   0&     |c_{1}(t)|^{2}&     c_{1}(t)c_{2}^{*}(t)&                             0  \\
   0&  c_{1}^{*}(t)c_{2}(t)&        |c_{2}(t)|^{2}&                             0  \\
   0&               0&                  0&      1-|c_{1}(t)|^{2}-|c_{2}(t)|^{2}  \\
  \end{matrix}
  \right).
\end{equation}
The dynamics of the qubit-magnon entanglement can be quantified by the concurrence $C=2|c_{1}(t)c_{2}^{*}(t)|$ \cite{Maniscalco2008}, which
ranges from 0 (separable state) to 1 (maximally entangled one).
Fig. 5 displays the numerical results of the time evolution
of the concurrence and populations of the qubit and the magnon mode in the presence of decoherence.
Obviously, only when the magnetic sphere is placed to the left-side cavity $n=-1$,
the qubit-magnon entanglement can occur.

\subsection{Chiral magnon blockade}

Second, we discuss how to realize a directional magnon blockade via the engineered chiral qubit-magnon coupling. Magnon blockade describes the process that a magnon mode absorbing the first magnon will block subsequent ones. In our scheme, the chiral magnon blockade relies on the strong anharmonicity of dressed states of the chiral coupled qubit-magnon system.

To generate the directional magnon blockade, we apply a microwave field
$\mathcal{H}_{p}=\xi(m^{\dag}e^{-i\omega_{p}{t}}+me^{i\omega_{p}{t}})$
with the amplitude $\xi$ and the frequency $\omega_{p}$ to directly drive the magnetic sphere.
When the magnetic sphere is loaded on the left-side cavity $n=-1$,
the qubit-magnon interaction is governed by the Hamiltonian $\mathcal{H}_{e}$.
In the limit $\xi\ll{G}$, we can diagonalize $\mathcal{H}_{e}$,
and obtain the eigenvalues $E_{\pm}=M\omega_{m}^{\prime}\pm\sqrt{M}G$ ($M\geq1$)
and the dressed states $|M,\pm\rangle=(|M-1,e\rangle\mp|M,g\rangle)/\sqrt{2}$.
As shown in Fig. 6(a), when the driving field resonates with the transitions $|0,g\rangle\leftrightarrow|1,\pm\rangle$, it will be highly
detuned from other higher-order transitions under the condition $G\gg\{\gamma_{q},\gamma_{m}\}$. As a result, the presence of one magnon in the system will inhibit further magnon absorption, i.e., this is the mechanism of magnon blockade.
However, when the magnetic sphere is loaded on the right-side cavity $n=1$, the magnon mode is decoupled from the qubit.
In this case, the dressed states $|M,\pm\rangle$ are degenerate [see Fig. 6(b)], so there is no anharmonicity to create the blockade effect. Therefore, we can generate a chiral magnon blockade.

\begin{figure}[tbph]
\centerline{\includegraphics[width=8cm]{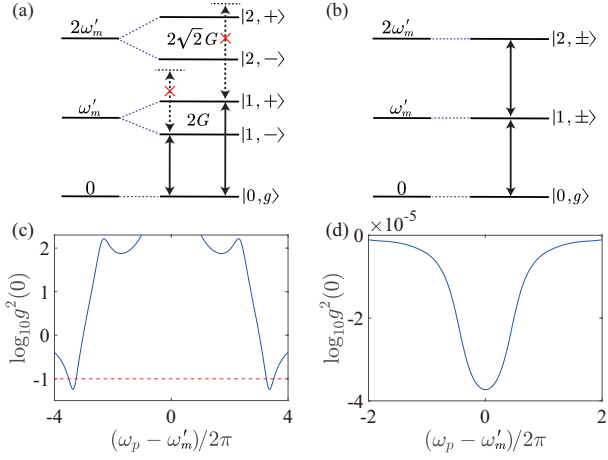}}
\caption{Energy-level diagram of the dressed states of the coupled qubit-magnon system,
where the magnetic sphere is placed to the (a) left-side or (b) right-side cavity.
(c,d) The steady-state logarithmic equal-time second-order
correlation function $\mathrm{log}_{10}g^{2}(0)$ as a function
of the detuning $(\omega_{p}-\omega_{m}^{\prime})/2\pi$ for $\xi/2\pi=0.03$ MHz.
[(c) $n=-1$; (d) $n=1$]. Other parameters are the same as in Fig. 5.}
\end{figure}

To reveal the statistical properties of the magnons, we introduce the equal-time second-order correlation function
$g^{(2)}(0)=\frac{\langle{m^{\dag}m^{\dag}mm}\rangle}{\langle{m^{\dag}m}\rangle^{2}}$ \cite{H8,H9},
which can be numerically calculated from the master equation
\begin{equation}
\begin{aligned}
\dot{\rho^{\prime}}=&-i[\mathcal{H}_{e}+\mathcal{H}_{p},\rho^{\prime}]
+\gamma_{q}\mathscr{L}[\sigma_{-}]\rho^{\prime}+\gamma_{m}\mathscr{L}[m]\rho^{\prime}.
\end{aligned}
\end{equation}
Here $g^{(2)}(0)<1$ represents the magnon antibunching \cite{E8,E9}.
In Fig. 6(c) and Fig. 6(d), we plot the steady-state logarithmic second-order correlation function
$\mathrm{log}_{10}g^{2}(0)$ as a function of the detuning $(\omega_{p}-\omega_{m}^{\prime})/2\pi$.
For the case of $n=-1$, $\mathrm{log}_{10}g^{2}(0)\approx-1.24$ ($g^{2}(0)\approx0.057$)
is obtained with $\omega_{p}=\omega_{m}^{\prime}\pm{G}$, implying the strong magnon antibunching effect.
In stark contrast, there is almost no magnon antibunching effect for $n=1$.

\section{Conclusion}

In conclusion, we have presented a novel strategy to achieve a chiral coupling between a magnon mode and a superconducting qubit mediated by a coupled-cavity array. Based on the engineered chiral interaction, we demonstrated the chiral qubit-magnon entanglement and the chiral magnon blockade. The present work paves an appealing way for creating chirality in magnonics systems, and is expected to stimulate a series of quantum technological applications, such as one-way preparation of magnon nonclassical states, chiral magnon-induced transparency, and chiral light-microwave conversion.

\section{Acknowledgements}
The work was supported by the National Nature Science Foundation of China (Grant Nos. 11704306 and 12074307).

\begin{widetext}
\subsection*{Appendix A: Total Hamiltonian}
In this section, we illustrate the detailed derivation of the total Hamiltonian in Eq. (4) of the main text.
By performing a unitary transformation
$U_{1}(t)=\mathrm{exp}[(\sum_{k}\omega_{k}a^{\dag}_{k}a_{k}+\omega_{m}m^{\dag}m+\omega_{q}\sigma_{+}\sigma_{-})t]$,
we have
\begin{equation}\tag{A1}
\begin{aligned}
\mathcal{H}_{I}=&\sum_{k}\Big\{\frac{g_{m}e^{ikn}}{\sqrt{N}}e^{i(\omega_{k}-\omega_{m})t}a_{k}^{\dag}m
+\frac{g_{q_{-l}}e^{-ikl}}{\sqrt{N}}e^{i[\omega_{k}-(\omega_{q}-\omega_{c})]t}e^{-i\theta_{-l}}a_{k}^{\dag}\sigma_{-}
+\frac{g_{q_{-l}}e^{-ikl}}{\sqrt{N}}e^{i[\omega_{k}-(\omega_{q}+\omega_{c})]t}e^{i\theta_{-l}}a_{k}^{\dag}\sigma_{-}\\
&+\frac{g_{q_{l}}e^{ikl}}{\sqrt{N}}e^{i[\omega_{k}-(\omega_{q}-\omega_{c})]t}e^{-i\theta_{l}}a_{k}^{\dag}\sigma_{-}
+\frac{g_{q_{l}}e^{ikl}}{\sqrt{N}}e^{i[\omega_{k}-(\omega_{q}+\omega_{c})]t}e^{i\theta_{l}}a_{k}^{\dag}\sigma_{-}+\mathrm{H.c.}\Big\}.
\end{aligned}
\end{equation}
Obviously, we can find two side-band terms $\omega_{q}\pm\omega_{c}$ induced by the time-dependent coupling.
Here, we consider that the qubit's frequency is far-detuned from the down band edge and satisfy $\omega_{db}-\omega_{q}\gg\{g_{m},g_{q_{j}}\}$.
By choosing suitable $\omega_{c}$, the blue-side band $\omega_{q}+\omega_{c}$  is near the down band edge
and the red-side band $\omega_{q}-\omega_{c}$ is far away from the down band edge.
In our scheme, the frequency of the magnon mode can be tuned to meet $\omega_{m}=\omega_{q}+\omega_{c}$ by adjusting the external magnetic
filed. Therefore, those fast oscillating terms
\{$\frac{g_{q_{-l}}e^{-ikl}}{\sqrt{N}}e^{i[\omega_{k}-(\omega_{q}-\omega_{c})]t}e^{-i\theta_{-l}}a_{k}^{\dag}\sigma_{-}$,
$\frac{g_{q_{-l}}e^{ikl}}{\sqrt{N}}e^{-i[\omega_{k}-(\omega_{q}-\omega_{c})]t}e^{i\theta_{-l}}a_{k}\sigma_{+}$,
$\frac{g_{q_{l}}e^{ikl}}{\sqrt{N}}e^{i[\omega_{k}-(\omega_{q}-\omega_{c})]t}e^{-i\theta_{l}}a_{k}^{\dag}\sigma_{-}$,
$\frac{g_{q_{l}}e^{-ikl}}{\sqrt{N}}e^{-i[\omega_{k}-(\omega_{q}-\omega_{c})]t}e^{i\theta_{l}}a_{k}\sigma_{+}$\}
can be safely dropped under the rotating wave approximation.
As a result, the interaction picture Hamiltonian can be simplified as
\begin{equation}\tag{A2}
\begin{aligned}
\mathcal{H}_{I}&=\sum_{k}\Big\{\frac{g_{m}e^{ikn}}{\sqrt{N}}e^{i(\omega_{k}-\omega_{m})t}a_{k}^{\dag}m
+\frac{g_{q_{-l}}e^{-ikl}}{\sqrt{N}}e^{i(\omega_{k}-\omega_{m})t+i\theta_{-l}}a_{k}^{\dag}\sigma_{-}
+\frac{g_{q_{l}}e^{ikl}}{\sqrt{N}}e^{i(\omega_{k}-\omega_{m})t+i\theta_{l}}a_{k}^{\dag}\sigma_{-}+\mathrm{H.c.}\Big\}.
\end{aligned}
\end{equation}
In the frame rotated by the unitary transformation
$U_{2}(t)=\mathrm{exp}[-(\sum_{k}\omega_{k}a^{\dag}_{k}a_{k}+\omega_{m}m^{\dag}m+\omega_{m}\sigma_{+}\sigma_{-})t]$, we can obtain
\begin{equation}\tag{A3}
\begin{aligned}
\mathcal{H}&=\sum_{k}\omega_{k}a^{\dag}_{k}a_{k}+\omega_{m}m^{\dag}m+\omega_{m}\sigma_{+}\sigma_{-}
+\sum_{k}\Big[\frac{g_{m}e^{ikn}}{\sqrt{N}}a_{k}^{\dag}m
+\frac{g_{q_{-l}}e^{i(\theta_{-l}-kl)}+g_{q_{l}}e^{i(\theta_{l}+kl)}}{\sqrt{N}}a_{k}^{\dag}\sigma_{-}+\mathrm{H.c.}\Big].
\end{aligned}
\end{equation}

\subsection*{Appendix B: Chiral qubit-magnon interaction}
In this section, we present the procedures for the derivation of the effective Hamiltonian $\mathcal{H}_{e}$ in detail.
Let us now divide the total Hamiltonian $\mathcal{H}_{e}$ into two parts $\mathcal{H}_{e}=\mathcal{H}_{free}+\mathcal{H}_{int}$,
where the free part is
\begin{equation}\tag{B1}
\mathcal{H}_{free}=\sum_{k}\omega_{k}a^{\dag}_{k}a_{k}+\omega_{m}m^{\dag}m+\omega_{m}\sigma_{+}\sigma_{-},
\end{equation}
and the interaction part is
\begin{equation}\tag{B2}
\mathcal{H}_{int}=\sum_{k}\Big[\frac{g_{m}e^{ikn}}{\sqrt{N}}a_{k}^{\dag}m+\frac{g_{m}e^{-ikn}}{\sqrt{N}}a_{k}m^{\dag}
+\frac{g_{q_{-l}}e^{i(\theta_{-l}-kl)}+g_{q_{l}}e^{i(\theta_{l}+kl)}}{\sqrt{N}}a_{k}^{\dag}\sigma_{-}
+\frac{g_{q_{-l}}e^{-i(\theta_{-l}-kl)}+g_{q_{l}}e^{-i(\theta_{l}+kl)}}{\sqrt{N}}a_{k}\sigma_{+}\Big].
\end{equation}
In the dispersive regime, the down band edge satisfies $\omega_{k}-\omega_{m}\gg{g_{q_{j}}/\sqrt{N}}$ and $\omega_{k}-\omega_{m}\gg{g_{m}/\sqrt{N}}$.
In this case, we can apply the Frohlich-Nakajima transformation.
It needs to find a unitary transformation $\mathcal{U}=\mathrm{exp}(\mathcal{-V})$,
where $\mathcal{V}$ is an anti-Hermitian operator $\mathcal{V}^{\dag}=-\mathcal{V}$ and meets $\mathcal{H}_{int}+[\mathcal{H}_{free},\mathcal{V}]=0$.
If we take $\mathcal{V}=\sum_{k}[(M_{k}a_{k}m^{\dag}-M_{k}^{*}a_{k}^{\dag}m)+(Q_{k}a_{k}\sigma_{+}-Q_{k}^{*}a_{k}^{\dag}\sigma_{-})]$, we have
\begin{equation}\tag{B3}
\begin{aligned}
&\mathcal{H}_{int}+[\mathcal{H}_{free},\mathcal{V}] \\
=&[\frac{g_{m}e^{ikn}}{\sqrt{N}}a_{k}^{\dag}m+\frac{g_{m}e^{-ikn}}{\sqrt{N}}a_{k}m^{\dag}
+\frac{g_{q_{-l}}e^{i(\theta_{-l}-kl)}+g_{q_{l}}e^{i(\theta_{l}+kl)}}{\sqrt{N}}a_{k}^{\dag}\sigma_{-}
+\frac{g_{q_{-l}}e^{-i(\theta_{-l}-kl)}+g_{q_{l}}e^{-i(\theta_{l}+kl)}}{\sqrt{N}}a_{k}\sigma_{+}\Big] \\
&-\sum_{k}(\omega_{k}-\omega_{m})(M_{k}a_{k}m^{\dag}+M_{k}^{*}a_{k}^{\dag}m)
-\sum_{k}(\omega_{k}-\omega_{m})(Q_{k}a_{k}\sigma_{+}+Q_{k}^{*}a_{k}^{\dag}\sigma_{-})=0.
\end{aligned}
\end{equation}
So, we can obtain $M_{k}=\frac{g_{m}e^{-ikn}}{\sqrt{N}(\omega_{k}-\omega_{m})}$
and $Q_{k}=\frac{g_{q_{-l}}e^{-i(\theta_{-l}-kl)}+g_{q_{l}}e^{-i(\theta_{l}+kl)}}{\sqrt{N}(\omega_{k}-\omega_{m})}$.
Because the coefficients $M_{k}$ and $Q_{k}$ are small in the large detuning regime,
the higher-order terms can be dropped and only the second-order term $[\mathcal{H}_{int},\mathcal{V}]$ should be taken into account.
Therefore, the effective Hamiltonian can be derived as
\begin{equation}\tag{B4}
\begin{aligned}
&\mathcal{H}_{e}=\mathcal{U}\mathcal{H}\mathcal{U}^{\dag}\simeq{\mathcal{H}_{free}+\frac{1}{2}[\mathcal{H}_{int},\mathcal{V}]} \\
=&\sum_{k}\omega_{k}a^{\dag}_{k}a_{k}
+\frac{1}{N}\sum_{k}\sum_{k^{\prime}}\frac{g_{m}^{2}}{\omega_{k^{\prime}}-\omega_{m}}a_{k}^{\dag}a_{k^{\prime}}
-\frac{1}{N}\sum_{k}\sum_{k^{\prime}}\frac{[g_{q_{-l}}e^{i(\theta_{-l}-kl)}+g_{q_{l}}e^{i(\theta_{l}+kl)}][g_{q_{-l}}e^{-i(\theta_{-l}-kl)}+g_{q_{l}}e^{-i(\theta_{l}+kl)}]}
{\omega_{k^{\prime}}-\omega_{m}}a_{k}^{\dag}a_{k^{\prime}}\sigma_{z} \\
&+\frac{1}{N}\sum_{k}(\omega_{m}-\frac{g_{m}^{2}}{\omega_{k}-\omega_{m}})m^{\dag}m
+\frac{1}{N}\sum_{k}\Big\{\omega_{m}-\frac{[g_{q_{-l}}e^{i(\theta_{-l}-kl)}+g_{q_{l}}e^{i(\theta_{l}+kl)}][g_{q_{-l}}e^{-i(\theta_{-l}-kl)}+g_{q_{l}}e^{-i(\theta_{l}+kl)}]}
{\omega_{k}-\omega_{m}}\Big\}\sigma_{+}\sigma_{-}\\
&-\frac{1}{N}\sum_{k}\frac{g_{q_{-l}}g_{m}e^{i(\theta_{-l}-kl)}e^{-ikn}+g_{q_{l}}g_{m}e^{i(\theta_{l}+kl)}e^{-ikn}}{\omega_{k}-\omega_{m}}m^{\dag}\sigma_{-}
-\frac{1}{N}\sum_{k}\frac{g_{q_{-l}}g_{m}e^{-i(\theta_{-l}-kl)}e^{ikn}+g_{q_{l}}g_{m}e^{-i(\theta_{l}+kl)}e^{ikn}}{\omega_{k}-\omega_{m}}m\sigma_{+}.\\
\end{aligned}
\end{equation}
In the dispersive regime, the photonic modes of the coupled-resonator array are in the vacuum state.
Substituting $\omega_{k}=\omega_{r}+2J\mathrm{cos}(k)$ into Eq. (B4),
the effective Hamiltonian can be rewritten as
\begin{equation}\tag{B5}
\mathcal{H}_{e}=\omega_{m}^{\prime}m^{\dag}m+\omega_{q}^{\prime}\sigma_{+}\sigma_{-}-Gm^{\dag}\sigma_{-}-G^{*}m\sigma_{+},
\end{equation}
with
\begin{equation}\tag{B6}
\begin{aligned}
\omega_{m}^{\prime}&=\omega_{m}-\int_{-\pi}^{\pi}\frac{g_{m}^{2}}{\omega_{r}-\omega_{m}+2J\mathrm{cos}(k)}\frac{dk}{2\pi} \\
&=\omega_{m}-\frac{g_{m}^{2}}{\sqrt{(\omega_{db}-\omega_{m})^{2}+4J(\omega_{db}-\omega_{m})}}, \\
\omega_{q}^{\prime}&=\omega_{m}-\int_{-\pi}^{\pi}\frac{[g_{q_{-l}}e^{i(\theta_{-l}-kl)}+g_{q_{l}}e^{i(\theta_{l}+kl)}][g_{q_{-l}}e^{-i(\theta_{-l}-kl)}+g_{q_{l}}e^{-i(\theta_{l}+kl)}]}
{\omega_{r}-\omega_{m}+2J\mathrm{cos}(k)}\frac{dk}{2\pi}\\
&=\omega_{m}-\frac{g_{q_{-l}}^{2}+g_{q_{l}}^{2}+2\mathrm{cos}(\theta_{-l}-\theta_{l})g_{q_{l}}g_{q_{-l}}e^{-|2l|[\mathrm{arccosh}(1+(\omega_{db}-\omega_{m})/2J]}}
{\sqrt{(\omega_{db}-\omega_{m})^{2}+4J(\omega_{db}-\omega_{m})}}, \\
G&=\int_{-\pi}^{\pi}\frac{g_{q_{-l}}g_{m}e^{i(\theta_{-l}-kl)}e^{-ikn}+g_{q_{l}}g_{m}e^{i(\theta_{l}+kl)}e^{-ikn}}
{\omega_{r}-\omega_{m}+2J\mathrm{cos}(k)}\frac{dk}{2\pi} \\
&=\frac{(-1)^{|n+l|}g_{q_{-l}}g_{m}e^{i\theta_{-l}}e^{-|n+l|[\mathrm{arccosh}(1+(\omega_{db}-\omega_{m})/2J]}
+(-1)^{|n-l|}g_{q_{l}}g_{m}e^{i\theta_{l}}e^{-|n-l|[\mathrm{arccosh}(1+(\omega_{db}-\omega_{m})/2J]}}
{\sqrt{(\omega_{db}-\omega_{m})^{2}+4J(\omega_{db}-\omega_{m})}},
\end{aligned}
\end{equation}
where we have replaced the discrete modes by the continuous distribution, i.e., $\frac{1}{N}\sum_{k}\rightarrow\int_{-\pi}^{\pi}\frac{dk}{2\pi}$.

\end{widetext}

\bibliography{Ref}

\begin{thebibliography}{92}%
\makeatletter
\providecommand \@ifxundefined [1]{%
 \@ifx{#1\undefined}
}%
\providecommand \@ifnum [1]{%
 \ifnum #1\expandafter \@firstoftwo
 \else \expandafter \@secondoftwo
 \fi
}%
\providecommand \@ifx [1]{%
 \ifx #1\expandafter \@firstoftwo
 \else \expandafter \@secondoftwo
 \fi
}%
\providecommand \natexlab [1]{#1}%
\providecommand \enquote  [1]{``#1''}%
\providecommand \bibnamefont  [1]{#1}%
\providecommand \bibfnamefont [1]{#1}%
\providecommand \citenamefont [1]{#1}%
\providecommand \href@noop [0]{\@secondoftwo}%
\providecommand \href [0]{\begingroup \@sanitize@url \@href}%
\providecommand \@href[1]{\@@startlink{#1}\@@href}%
\providecommand \@@href[1]{\endgroup#1\@@endlink}%
\providecommand \@sanitize@url [0]{\catcode `\\12\catcode `\$12\catcode
  `\&12\catcode `\#12\catcode `\^12\catcode `\_12\catcode `\%12\relax}%
\providecommand \@@startlink[1]{}%
\providecommand \@@endlink[0]{}%
\providecommand \url  [0]{\begingroup\@sanitize@url \@url }%
\providecommand \@url [1]{\endgroup\@href {#1}{\urlprefix }}%
\providecommand \urlprefix  [0]{URL }%
\providecommand \Eprint [0]{\href }%
\providecommand \doibase [0]{http://dx.doi.org/}%
\providecommand \selectlanguage [0]{\@gobble}%
\providecommand \bibinfo  [0]{\@secondoftwo}%
\providecommand \bibfield  [0]{\@secondoftwo}%
\providecommand \translation [1]{[#1]}%
\providecommand \BibitemOpen [0]{}%
\providecommand \bibitemStop [0]{}%
\providecommand \bibitemNoStop [0]{.\EOS\space}%
\providecommand \EOS [0]{\spacefactor3000\relax}%
\providecommand \BibitemShut  [1]{\csname bibitem#1\endcsname}%
\let\auto@bib@innerbib\@empty
\bibitem [{\citenamefont {Pichler}\ \emph {et~al.}(2015)\citenamefont
  {Pichler}, \citenamefont {Ramos}, \citenamefont {Daley},\ and\ \citenamefont
  {Zoller}}]{FA1}%
  \BibitemOpen
  \bibfield  {author} {\bibinfo {author} {\bibfnamefont {H.}~\bibnamefont
  {Pichler}}, \bibinfo {author} {\bibfnamefont {T.}~\bibnamefont {Ramos}},
  \bibinfo {author} {\bibfnamefont {A.~J.}\ \bibnamefont {Daley}}, \ and\
  \bibinfo {author} {\bibfnamefont {P.}~\bibnamefont {Zoller}},\ }\href
  {\doibase 10.1103/PhysRevA.91.042116} {\bibfield  {journal} {\bibinfo
  {journal} {Phys. Rev. A}\ }\textbf {\bibinfo {volume} {91}},\ \bibinfo
  {pages} {042116} (\bibinfo {year} {2015})}\BibitemShut {NoStop}%
\bibitem [{\citenamefont {Xi}\ \emph {et~al.}(2021)\citenamefont {Xi},
  \citenamefont {Ma}, \citenamefont {Wan}, \citenamefont {Dong},\ and\
  \citenamefont {Sun}}]{Xi2021}%
  \BibitemOpen
  \bibfield  {author} {\bibinfo {author} {\bibfnamefont {X.}~\bibnamefont
  {Xi}}, \bibinfo {author} {\bibfnamefont {J.}~\bibnamefont {Ma}}, \bibinfo
  {author} {\bibfnamefont {S.}~\bibnamefont {Wan}}, \bibinfo {author}
  {\bibfnamefont {C.-H.}\ \bibnamefont {Dong}}, \ and\ \bibinfo {author}
  {\bibfnamefont {X.}~\bibnamefont {Sun}},\ }\href {\doibase
  10.1126/sciadv.abe1398} {\bibfield  {journal} {\bibinfo  {journal} {Science
  Advances}\ }\textbf {\bibinfo {volume} {7}} (\bibinfo {year} {2021}),\
  10.1126/sciadv.abe1398}\BibitemShut {NoStop}%
\bibitem [{\citenamefont {Owens}\ \emph {et~al.}(2022)\citenamefont {Owens},
  \citenamefont {Panetta}, \citenamefont {Saxberg}, \citenamefont {Roberts},
  \citenamefont {Chakram}, \citenamefont {Ma}, \citenamefont {Vrajitoarea},
  \citenamefont {Simon},\ and\ \citenamefont {Schuster}}]{Owens2022}%
  \BibitemOpen
  \bibfield  {author} {\bibinfo {author} {\bibfnamefont {J.~C.}\ \bibnamefont
  {Owens}}, \bibinfo {author} {\bibfnamefont {M.~G.}\ \bibnamefont {Panetta}},
  \bibinfo {author} {\bibfnamefont {B.}~\bibnamefont {Saxberg}}, \bibinfo
  {author} {\bibfnamefont {G.}~\bibnamefont {Roberts}}, \bibinfo {author}
  {\bibfnamefont {S.}~\bibnamefont {Chakram}}, \bibinfo {author} {\bibfnamefont
  {R.}~\bibnamefont {Ma}}, \bibinfo {author} {\bibfnamefont {A.}~\bibnamefont
  {Vrajitoarea}}, \bibinfo {author} {\bibfnamefont {J.}~\bibnamefont {Simon}},
  \ and\ \bibinfo {author} {\bibfnamefont {D.~I.}\ \bibnamefont {Schuster}},\
  }\href {\doibase 10.1038/s41567-022-01671-3} {\bibfield  {journal} {\bibinfo
  {journal} {Nature Physics}\ }\textbf {\bibinfo {volume} {18}},\ \bibinfo
  {pages} {1048} (\bibinfo {year} {2022})}\BibitemShut {NoStop}%
\bibitem [{\citenamefont {Nechayev}\ and\ \citenamefont {Banzer}(2019)}]{F1}%
  \BibitemOpen
  \bibfield  {author} {\bibinfo {author} {\bibfnamefont {S.}~\bibnamefont
  {Nechayev}}\ and\ \bibinfo {author} {\bibfnamefont {P.}~\bibnamefont
  {Banzer}},\ }\href {\doibase 10.1103/physrevb.99.241101} {\bibfield
  {journal} {\bibinfo  {journal} {Physical Review B}\ }\textbf {\bibinfo
  {volume} {99}},\ \bibinfo {pages} {241101} (\bibinfo {year}
  {2019})}\BibitemShut {NoStop}%
\bibitem [{\citenamefont {Yu}\ \emph {et~al.}(2022)\citenamefont {Yu},
  \citenamefont {Luo},\ and\ \citenamefont {Bauer}}]{YuT2022}%
  \BibitemOpen
  \bibfield  {author} {\bibinfo {author} {\bibfnamefont {T.}~\bibnamefont
  {Yu}}, \bibinfo {author} {\bibfnamefont {Z.}~\bibnamefont {Luo}}, \ and\
  \bibinfo {author} {\bibfnamefont {G.~E.~W.}\ \bibnamefont {Bauer}},\
  }\href@noop {} {\  (\bibinfo {year} {2022})},\ \Eprint
  {http://arxiv.org/abs/2206.05535} {arXiv:2206.05535 [cond-mat.mes-hall]}
  \BibitemShut {NoStop}%
\bibitem [{\citenamefont {Naaman}\ \emph {et~al.}(2019)\citenamefont {Naaman},
  \citenamefont {Paltiel},\ and\ \citenamefont {Waldeck}}]{NA2019}%
  \BibitemOpen
  \bibfield  {author} {\bibinfo {author} {\bibfnamefont {R.}~\bibnamefont
  {Naaman}}, \bibinfo {author} {\bibfnamefont {Y.}~\bibnamefont {Paltiel}}, \
  and\ \bibinfo {author} {\bibfnamefont {D.~H.}\ \bibnamefont {Waldeck}},\
  }\href {\doibase 10.1038/s41570-019-0087-1} {\bibfield  {journal} {\bibinfo
  {journal} {Nature Reviews Chemistry}\ }\textbf {\bibinfo {volume} {3}},\
  \bibinfo {pages} {250} (\bibinfo {year} {2019})}\BibitemShut {NoStop}%
\bibitem [{\citenamefont {Döring}\ \emph {et~al.}(2022)\citenamefont
  {Döring}, \citenamefont {Ushakova},\ and\ \citenamefont {Rogach}}]{DR2022}%
  \BibitemOpen
  \bibfield  {author} {\bibinfo {author} {\bibfnamefont {A.}~\bibnamefont
  {Döring}}, \bibinfo {author} {\bibfnamefont {E.}~\bibnamefont {Ushakova}}, \
  and\ \bibinfo {author} {\bibfnamefont {A.~L.}\ \bibnamefont {Rogach}},\
  }\href {\doibase 10.1038/s41377-022-00764-1} {\bibfield  {journal} {\bibinfo
  {journal} {Light: Science {\&} Applications}\ }\textbf {\bibinfo {volume}
  {11}} (\bibinfo {year} {2022}),\ 10.1038/s41377-022-00764-1}\BibitemShut
  {NoStop}%
\bibitem [{\citenamefont {Das}\ \emph {et~al.}(2022)\citenamefont {Das},
  \citenamefont {Kundelev}, \citenamefont {Vedernikova}, \citenamefont
  {Cherevkov}, \citenamefont {Danilov}, \citenamefont {Koroleva}, \citenamefont
  {Zhizhin}, \citenamefont {Tsypkin}, \citenamefont {Litvin}, \citenamefont
  {Baranov}, \citenamefont {Fedorov}, \citenamefont {Ushakova},\ and\
  \citenamefont {Rogach}}]{Das2022}%
  \BibitemOpen
  \bibfield  {author} {\bibinfo {author} {\bibfnamefont {A.}~\bibnamefont
  {Das}}, \bibinfo {author} {\bibfnamefont {E.~V.}\ \bibnamefont {Kundelev}},
  \bibinfo {author} {\bibfnamefont {A.~A.}\ \bibnamefont {Vedernikova}},
  \bibinfo {author} {\bibfnamefont {S.~A.}\ \bibnamefont {Cherevkov}}, \bibinfo
  {author} {\bibfnamefont {D.~V.}\ \bibnamefont {Danilov}}, \bibinfo {author}
  {\bibfnamefont {A.~V.}\ \bibnamefont {Koroleva}}, \bibinfo {author}
  {\bibfnamefont {E.~V.}\ \bibnamefont {Zhizhin}}, \bibinfo {author}
  {\bibfnamefont {A.~N.}\ \bibnamefont {Tsypkin}}, \bibinfo {author}
  {\bibfnamefont {A.~P.}\ \bibnamefont {Litvin}}, \bibinfo {author}
  {\bibfnamefont {A.~V.}\ \bibnamefont {Baranov}}, \bibinfo {author}
  {\bibfnamefont {A.~V.}\ \bibnamefont {Fedorov}}, \bibinfo {author}
  {\bibfnamefont {E.~V.}\ \bibnamefont {Ushakova}}, \ and\ \bibinfo {author}
  {\bibfnamefont {A.~L.}\ \bibnamefont {Rogach}},\ }\href {\doibase
  10.1038/s41377-022-00778-9} {\bibfield  {journal} {\bibinfo  {journal}
  {Light: Science {\&} Applications}\ }\textbf {\bibinfo {volume} {11}}
  (\bibinfo {year} {2022}),\ 10.1038/s41377-022-00778-9}\BibitemShut {NoStop}%
\bibitem [{\citenamefont {Ramos}\ \emph {et~al.}(2016)\citenamefont {Ramos},
  \citenamefont {Vermersch}, \citenamefont {Hauke}, \citenamefont {Pichler},\
  and\ \citenamefont {Zoller}}]{Ramos2016}%
  \BibitemOpen
  \bibfield  {author} {\bibinfo {author} {\bibfnamefont {T.}~\bibnamefont
  {Ramos}}, \bibinfo {author} {\bibfnamefont {B.}~\bibnamefont {Vermersch}},
  \bibinfo {author} {\bibfnamefont {P.}~\bibnamefont {Hauke}}, \bibinfo
  {author} {\bibfnamefont {H.}~\bibnamefont {Pichler}}, \ and\ \bibinfo
  {author} {\bibfnamefont {P.}~\bibnamefont {Zoller}},\ }\href {\doibase
  10.1103/PhysRevA.93.062104} {\bibfield  {journal} {\bibinfo  {journal} {Phys.
  Rev. A}\ }\textbf {\bibinfo {volume} {93}},\ \bibinfo {pages} {062104}
  (\bibinfo {year} {2016})}\BibitemShut {NoStop}%
\bibitem [{\citenamefont {Coles}\ \emph {et~al.}(2016)\citenamefont {Coles},
  \citenamefont {Price}, \citenamefont {Dixon}, \citenamefont {Royall},
  \citenamefont {Clarke}, \citenamefont {Kok}, \citenamefont {Skolnick},
  \citenamefont {Fox},\ and\ \citenamefont {Makhonin}}]{Coles2016}%
  \BibitemOpen
  \bibfield  {author} {\bibinfo {author} {\bibfnamefont {R.~J.}\ \bibnamefont
  {Coles}}, \bibinfo {author} {\bibfnamefont {D.~M.}\ \bibnamefont {Price}},
  \bibinfo {author} {\bibfnamefont {J.~E.}\ \bibnamefont {Dixon}}, \bibinfo
  {author} {\bibfnamefont {B.}~\bibnamefont {Royall}}, \bibinfo {author}
  {\bibfnamefont {E.}~\bibnamefont {Clarke}}, \bibinfo {author} {\bibfnamefont
  {P.}~\bibnamefont {Kok}}, \bibinfo {author} {\bibfnamefont {M.~S.}\
  \bibnamefont {Skolnick}}, \bibinfo {author} {\bibfnamefont {A.~M.}\
  \bibnamefont {Fox}}, \ and\ \bibinfo {author} {\bibfnamefont {M.~N.}\
  \bibnamefont {Makhonin}},\ }\href {\doibase 10.1038/ncomms11183} {\bibfield
  {journal} {\bibinfo  {journal} {Nature Communications}\ }\textbf {\bibinfo
  {volume} {7}} (\bibinfo {year} {2016}),\ 10.1038/ncomms11183}\BibitemShut
  {NoStop}%
\bibitem [{\citenamefont {Tang}\ and\ \citenamefont {Cohen}(2010)}]{F4}%
  \BibitemOpen
  \bibfield  {author} {\bibinfo {author} {\bibfnamefont {Y.}~\bibnamefont
  {Tang}}\ and\ \bibinfo {author} {\bibfnamefont {A.~E.}\ \bibnamefont
  {Cohen}},\ }\href {\doibase 10.1103/physrevlett.104.163901} {\bibfield
  {journal} {\bibinfo  {journal} {Physical Review Letters}\ }\textbf {\bibinfo
  {volume} {104}},\ \bibinfo {pages} {163901} (\bibinfo {year}
  {2010})}\BibitemShut {NoStop}%
\bibitem [{\citenamefont {Yoo}\ and\ \citenamefont {Park}(2015)}]{F5}%
  \BibitemOpen
  \bibfield  {author} {\bibinfo {author} {\bibfnamefont {S.}~\bibnamefont
  {Yoo}}\ and\ \bibinfo {author} {\bibfnamefont {Q.-H.}\ \bibnamefont {Park}},\
  }\href {\doibase 10.1103/physrevlett.114.203003} {\bibfield  {journal}
  {\bibinfo  {journal} {Physical Review Letters}\ }\textbf {\bibinfo {volume}
  {114}},\ \bibinfo {pages} {203003} (\bibinfo {year} {2015})}\BibitemShut
  {NoStop}%
\bibitem [{\citenamefont {Söllner}\ \emph {et~al.}(2015)\citenamefont
  {Söllner}, \citenamefont {Mahmoodian}, \citenamefont {Hansen}, \citenamefont
  {Midolo}, \citenamefont {Javadi}, \citenamefont {Kir{\v{s}}ansk{\.{e}}},
  \citenamefont {Pregnolato}, \citenamefont {El-Ella}, \citenamefont {Lee},
  \citenamefont {Song}, \citenamefont {Stobbe},\ and\ \citenamefont
  {Lodahl}}]{Soellner2015}%
  \BibitemOpen
  \bibfield  {author} {\bibinfo {author} {\bibfnamefont {I.}~\bibnamefont
  {Söllner}}, \bibinfo {author} {\bibfnamefont {S.}~\bibnamefont
  {Mahmoodian}}, \bibinfo {author} {\bibfnamefont {S.~L.}\ \bibnamefont
  {Hansen}}, \bibinfo {author} {\bibfnamefont {L.}~\bibnamefont {Midolo}},
  \bibinfo {author} {\bibfnamefont {A.}~\bibnamefont {Javadi}}, \bibinfo
  {author} {\bibfnamefont {G.}~\bibnamefont {Kir{\v{s}}ansk{\.{e}}}}, \bibinfo
  {author} {\bibfnamefont {T.}~\bibnamefont {Pregnolato}}, \bibinfo {author}
  {\bibfnamefont {H.}~\bibnamefont {El-Ella}}, \bibinfo {author} {\bibfnamefont
  {E.~H.}\ \bibnamefont {Lee}}, \bibinfo {author} {\bibfnamefont {J.~D.}\
  \bibnamefont {Song}}, \bibinfo {author} {\bibfnamefont {S.}~\bibnamefont
  {Stobbe}}, \ and\ \bibinfo {author} {\bibfnamefont {P.}~\bibnamefont
  {Lodahl}},\ }\href {\doibase 10.1038/nnano.2015.159} {\bibfield  {journal}
  {\bibinfo  {journal} {Nature Nanotechnology}\ }\textbf {\bibinfo {volume}
  {10}},\ \bibinfo {pages} {775} (\bibinfo {year} {2015})}\BibitemShut
  {NoStop}%
\bibitem [{\citenamefont {Han}\ \emph {et~al.}(2019)\citenamefont {Han},
  \citenamefont {Lee}, \citenamefont {Hanke}, \citenamefont {Mokrousov},
  \citenamefont {Kim}, \citenamefont {Yoo}, \citenamefont {van Hees},
  \citenamefont {Kim}, \citenamefont {Lavrijsen}, \citenamefont {You},
  \citenamefont {Swagten}, \citenamefont {Jung},\ and\ \citenamefont
  {Kläui}}]{F6}%
  \BibitemOpen
  \bibfield  {author} {\bibinfo {author} {\bibfnamefont {D.-S.}\ \bibnamefont
  {Han}}, \bibinfo {author} {\bibfnamefont {K.}~\bibnamefont {Lee}}, \bibinfo
  {author} {\bibfnamefont {J.-P.}\ \bibnamefont {Hanke}}, \bibinfo {author}
  {\bibfnamefont {Y.}~\bibnamefont {Mokrousov}}, \bibinfo {author}
  {\bibfnamefont {K.-W.}\ \bibnamefont {Kim}}, \bibinfo {author} {\bibfnamefont
  {W.}~\bibnamefont {Yoo}}, \bibinfo {author} {\bibfnamefont {Y.~L.~W.}\
  \bibnamefont {van Hees}}, \bibinfo {author} {\bibfnamefont {T.-W.}\
  \bibnamefont {Kim}}, \bibinfo {author} {\bibfnamefont {R.}~\bibnamefont
  {Lavrijsen}}, \bibinfo {author} {\bibfnamefont {C.-Y.}\ \bibnamefont {You}},
  \bibinfo {author} {\bibfnamefont {H.~J.~M.}\ \bibnamefont {Swagten}},
  \bibinfo {author} {\bibfnamefont {M.-H.}\ \bibnamefont {Jung}}, \ and\
  \bibinfo {author} {\bibfnamefont {M.}~\bibnamefont {Kläui}},\ }\href
  {\doibase 10.1038/s41563-019-0370-z} {\bibfield  {journal} {\bibinfo
  {journal} {Nature Materials}\ }\textbf {\bibinfo {volume} {18}},\ \bibinfo
  {pages} {703} (\bibinfo {year} {2019})}\BibitemShut {NoStop}%
\bibitem [{\citenamefont {Lodahl}\ \emph {et~al.}(2017)\citenamefont {Lodahl},
  \citenamefont {Mahmoodian}, \citenamefont {Stobbe}, \citenamefont
  {Rauschenbeutel}, \citenamefont {Schneeweiss}, \citenamefont {Volz},
  \citenamefont {Pichler},\ and\ \citenamefont {Zoller}}]{F2}%
  \BibitemOpen
  \bibfield  {author} {\bibinfo {author} {\bibfnamefont {P.}~\bibnamefont
  {Lodahl}}, \bibinfo {author} {\bibfnamefont {S.}~\bibnamefont {Mahmoodian}},
  \bibinfo {author} {\bibfnamefont {S.}~\bibnamefont {Stobbe}}, \bibinfo
  {author} {\bibfnamefont {A.}~\bibnamefont {Rauschenbeutel}}, \bibinfo
  {author} {\bibfnamefont {P.}~\bibnamefont {Schneeweiss}}, \bibinfo {author}
  {\bibfnamefont {J.}~\bibnamefont {Volz}}, \bibinfo {author} {\bibfnamefont
  {H.}~\bibnamefont {Pichler}}, \ and\ \bibinfo {author} {\bibfnamefont
  {P.}~\bibnamefont {Zoller}},\ }\href {\doibase 10.1038/nature21037}
  {\bibfield  {journal} {\bibinfo  {journal} {Nature}\ }\textbf {\bibinfo
  {volume} {541}},\ \bibinfo {pages} {473} (\bibinfo {year}
  {2017})}\BibitemShut {NoStop}%
\bibitem [{\citenamefont {Roushan}\ \emph {et~al.}(2016)\citenamefont {Roushan}
  \emph {et~al.}}]{Roushan2016}%
  \BibitemOpen
  \bibfield  {author} {\bibinfo {author} {\bibfnamefont {P.}~\bibnamefont
  {Roushan}} \emph {et~al.},\ }\href {\doibase 10.1038/nphys3930} {\bibfield
  {journal} {\bibinfo  {journal} {Nature Physics}\ }\textbf {\bibinfo {volume}
  {13}},\ \bibinfo {pages} {146} (\bibinfo {year} {2016})}\BibitemShut
  {NoStop}%
\bibitem [{\citenamefont {Vermersch}\ \emph {et~al.}(2017)\citenamefont
  {Vermersch}, \citenamefont {Guimond}, \citenamefont {Pichler},\ and\
  \citenamefont {Zoller}}]{F10}%
  \BibitemOpen
  \bibfield  {author} {\bibinfo {author} {\bibfnamefont {B.}~\bibnamefont
  {Vermersch}}, \bibinfo {author} {\bibfnamefont {P.-O.}\ \bibnamefont
  {Guimond}}, \bibinfo {author} {\bibfnamefont {H.}~\bibnamefont {Pichler}}, \
  and\ \bibinfo {author} {\bibfnamefont {P.}~\bibnamefont {Zoller}},\ }\href
  {\doibase 10.1103/physrevlett.118.133601} {\bibfield  {journal} {\bibinfo
  {journal} {Physical Review Letters}\ }\textbf {\bibinfo {volume} {118}},\
  \bibinfo {pages} {133601} (\bibinfo {year} {2017})}\BibitemShut {NoStop}%
\bibitem [{\citenamefont {Guimond}\ \emph {et~al.}(2020)\citenamefont
  {Guimond}, \citenamefont {Vermersch}, \citenamefont {Juan}, \citenamefont
  {Sharafiev}, \citenamefont {Kirchmair},\ and\ \citenamefont {Zoller}}]{F11}%
  \BibitemOpen
  \bibfield  {author} {\bibinfo {author} {\bibfnamefont {P.-O.}\ \bibnamefont
  {Guimond}}, \bibinfo {author} {\bibfnamefont {B.}~\bibnamefont {Vermersch}},
  \bibinfo {author} {\bibfnamefont {M.~L.}\ \bibnamefont {Juan}}, \bibinfo
  {author} {\bibfnamefont {A.}~\bibnamefont {Sharafiev}}, \bibinfo {author}
  {\bibfnamefont {G.}~\bibnamefont {Kirchmair}}, \ and\ \bibinfo {author}
  {\bibfnamefont {P.}~\bibnamefont {Zoller}},\ }\href {\doibase
  10.1038/s41534-020-0261-9} {\bibfield  {journal} {\bibinfo  {journal} {npj
  Quantum Information}\ }\textbf {\bibinfo {volume} {6}} (\bibinfo {year}
  {2020}),\ 10.1038/s41534-020-0261-9}\BibitemShut {NoStop}%
\bibitem [{\citenamefont {Hafezi}\ and\ \citenamefont {Rabl}(2012)}]{G1}%
  \BibitemOpen
  \bibfield  {author} {\bibinfo {author} {\bibfnamefont {M.}~\bibnamefont
  {Hafezi}}\ and\ \bibinfo {author} {\bibfnamefont {P.}~\bibnamefont {Rabl}},\
  }\href {\doibase 10.1364/oe.20.007672} {\bibfield  {journal} {\bibinfo
  {journal} {Optics Express}\ }\textbf {\bibinfo {volume} {20}},\ \bibinfo
  {pages} {7672} (\bibinfo {year} {2012})}\BibitemShut {NoStop}%
\bibitem [{\citenamefont {Shen}\ \emph {et~al.}(2016)\citenamefont {Shen},
  \citenamefont {Zhang}, \citenamefont {Chen}, \citenamefont {Zou},
  \citenamefont {Xiao}, \citenamefont {Zou}, \citenamefont {Sun}, \citenamefont
  {Guo},\ and\ \citenamefont {Dong}}]{G2}%
  \BibitemOpen
  \bibfield  {author} {\bibinfo {author} {\bibfnamefont {Z.}~\bibnamefont
  {Shen}}, \bibinfo {author} {\bibfnamefont {Y.-L.}\ \bibnamefont {Zhang}},
  \bibinfo {author} {\bibfnamefont {Y.}~\bibnamefont {Chen}}, \bibinfo {author}
  {\bibfnamefont {C.-L.}\ \bibnamefont {Zou}}, \bibinfo {author} {\bibfnamefont
  {Y.-F.}\ \bibnamefont {Xiao}}, \bibinfo {author} {\bibfnamefont {X.-B.}\
  \bibnamefont {Zou}}, \bibinfo {author} {\bibfnamefont {F.-W.}\ \bibnamefont
  {Sun}}, \bibinfo {author} {\bibfnamefont {G.-C.}\ \bibnamefont {Guo}}, \ and\
  \bibinfo {author} {\bibfnamefont {C.-H.}\ \bibnamefont {Dong}},\ }\href
  {\doibase 10.1038/nphoton.2016.161} {\bibfield  {journal} {\bibinfo
  {journal} {Nature Photonics}\ }\textbf {\bibinfo {volume} {10}},\ \bibinfo
  {pages} {657} (\bibinfo {year} {2016})}\BibitemShut {NoStop}%
\bibitem [{\citenamefont {Dong}\ \emph {et~al.}(2015)\citenamefont {Dong},
  \citenamefont {Shen}, \citenamefont {Zou}, \citenamefont {Zhang},
  \citenamefont {Fu},\ and\ \citenamefont {Guo}}]{G3}%
  \BibitemOpen
  \bibfield  {author} {\bibinfo {author} {\bibfnamefont {C.-H.}\ \bibnamefont
  {Dong}}, \bibinfo {author} {\bibfnamefont {Z.}~\bibnamefont {Shen}}, \bibinfo
  {author} {\bibfnamefont {C.-L.}\ \bibnamefont {Zou}}, \bibinfo {author}
  {\bibfnamefont {Y.-L.}\ \bibnamefont {Zhang}}, \bibinfo {author}
  {\bibfnamefont {W.}~\bibnamefont {Fu}}, \ and\ \bibinfo {author}
  {\bibfnamefont {G.-C.}\ \bibnamefont {Guo}},\ }\href {\doibase
  10.1038/ncomms7193} {\bibfield  {journal} {\bibinfo  {journal} {Nature
  Communications}\ }\textbf {\bibinfo {volume} {6}} (\bibinfo {year} {2015}),\
  10.1038/ncomms7193}\BibitemShut {NoStop}%
\bibitem [{\citenamefont {Kim}\ \emph {et~al.}(2015)\citenamefont {Kim},
  \citenamefont {Kuzyk}, \citenamefont {Han}, \citenamefont {Wang},\ and\
  \citenamefont {Bahl}}]{G4}%
  \BibitemOpen
  \bibfield  {author} {\bibinfo {author} {\bibfnamefont {J.}~\bibnamefont
  {Kim}}, \bibinfo {author} {\bibfnamefont {M.~C.}\ \bibnamefont {Kuzyk}},
  \bibinfo {author} {\bibfnamefont {K.}~\bibnamefont {Han}}, \bibinfo {author}
  {\bibfnamefont {H.}~\bibnamefont {Wang}}, \ and\ \bibinfo {author}
  {\bibfnamefont {G.}~\bibnamefont {Bahl}},\ }\href {\doibase
  10.1038/nphys3236} {\bibfield  {journal} {\bibinfo  {journal} {Nature
  Physics}\ }\textbf {\bibinfo {volume} {11}},\ \bibinfo {pages} {275}
  (\bibinfo {year} {2015})}\BibitemShut {NoStop}%
\bibitem [{\citenamefont {S{\'{a}}nchez-Burillo}\ \emph
  {et~al.}(2020)\citenamefont {S{\'{a}}nchez-Burillo}, \citenamefont {Wan},
  \citenamefont {Zueco},\ and\ \citenamefont {Gonz{\'{a}}lez-Tudela}}]{G5}%
  \BibitemOpen
  \bibfield  {author} {\bibinfo {author} {\bibfnamefont {E.}~\bibnamefont
  {S{\'{a}}nchez-Burillo}}, \bibinfo {author} {\bibfnamefont {C.}~\bibnamefont
  {Wan}}, \bibinfo {author} {\bibfnamefont {D.}~\bibnamefont {Zueco}}, \ and\
  \bibinfo {author} {\bibfnamefont {A.}~\bibnamefont {Gonz{\'{a}}lez-Tudela}},\
  }\href {\doibase 10.1103/physrevresearch.2.023003} {\bibfield  {journal}
  {\bibinfo  {journal} {Physical Review Research}\ }\textbf {\bibinfo {volume}
  {2}},\ \bibinfo {pages} {023003} (\bibinfo {year} {2020})}\BibitemShut
  {NoStop}%
\bibitem [{\citenamefont {De~Bernardis}\ \emph {et~al.}(2021)\citenamefont
  {De~Bernardis}, \citenamefont {Cian}, \citenamefont {Carusotto},
  \citenamefont {Hafezi},\ and\ \citenamefont {Rabl}}]{G6}%
  \BibitemOpen
  \bibfield  {author} {\bibinfo {author} {\bibfnamefont {D.}~\bibnamefont
  {De~Bernardis}}, \bibinfo {author} {\bibfnamefont {Z.-P.}\ \bibnamefont
  {Cian}}, \bibinfo {author} {\bibfnamefont {I.}~\bibnamefont {Carusotto}},
  \bibinfo {author} {\bibfnamefont {M.}~\bibnamefont {Hafezi}}, \ and\ \bibinfo
  {author} {\bibfnamefont {P.}~\bibnamefont {Rabl}},\ }\href {\doibase
  10.1103/PhysRevLett.126.103603} {\bibfield  {journal} {\bibinfo  {journal}
  {Phys. Rev. Lett.}\ }\textbf {\bibinfo {volume} {126}},\ \bibinfo {pages}
  {103603} (\bibinfo {year} {2021})}\BibitemShut {NoStop}%
\bibitem [{\citenamefont {Bello}\ \emph {et~al.}(2019)\citenamefont {Bello},
  \citenamefont {Platero}, \citenamefont {Cirac},\ and\ \citenamefont
  {Gonz{\'{a}}lez-Tudela}}]{G7}%
  \BibitemOpen
  \bibfield  {author} {\bibinfo {author} {\bibfnamefont {M.}~\bibnamefont
  {Bello}}, \bibinfo {author} {\bibfnamefont {G.}~\bibnamefont {Platero}},
  \bibinfo {author} {\bibfnamefont {J.~I.}\ \bibnamefont {Cirac}}, \ and\
  \bibinfo {author} {\bibfnamefont {A.}~\bibnamefont {Gonz{\'{a}}lez-Tudela}},\
  }\href {\doibase 10.1126/sciadv.aaw0297} {\bibfield  {journal} {\bibinfo
  {journal} {Science Advances}\ }\textbf {\bibinfo {volume} {5}} (\bibinfo
  {year} {2019}),\ 10.1126/sciadv.aaw0297}\BibitemShut {NoStop}%
\bibitem [{\citenamefont {Kim}\ \emph {et~al.}(2021)\citenamefont {Kim},
  \citenamefont {Zhang}, \citenamefont {Ferreira}, \citenamefont {Banker},
  \citenamefont {Iverson}, \citenamefont {Sipahigil}, \citenamefont {Bello},
  \citenamefont {Gonz{\'{a}}lez-Tudela}, \citenamefont {Mirhosseini},\ and\
  \citenamefont {Painter}}]{G8}%
  \BibitemOpen
  \bibfield  {author} {\bibinfo {author} {\bibfnamefont {E.}~\bibnamefont
  {Kim}}, \bibinfo {author} {\bibfnamefont {X.}~\bibnamefont {Zhang}}, \bibinfo
  {author} {\bibfnamefont {V.~S.}\ \bibnamefont {Ferreira}}, \bibinfo {author}
  {\bibfnamefont {J.}~\bibnamefont {Banker}}, \bibinfo {author} {\bibfnamefont
  {J.~K.}\ \bibnamefont {Iverson}}, \bibinfo {author} {\bibfnamefont
  {A.}~\bibnamefont {Sipahigil}}, \bibinfo {author} {\bibfnamefont
  {M.}~\bibnamefont {Bello}}, \bibinfo {author} {\bibfnamefont
  {A.}~\bibnamefont {Gonz{\'{a}}lez-Tudela}}, \bibinfo {author} {\bibfnamefont
  {M.}~\bibnamefont {Mirhosseini}}, \ and\ \bibinfo {author} {\bibfnamefont
  {O.}~\bibnamefont {Painter}},\ }\href {\doibase 10.1103/physrevx.11.011015}
  {\bibfield  {journal} {\bibinfo  {journal} {Physical Review X}\ }\textbf
  {\bibinfo {volume} {11}},\ \bibinfo {pages} {011015} (\bibinfo {year}
  {2021})}\BibitemShut {NoStop}%
\bibitem [{\citenamefont {Xu}\ \emph {et~al.}(2021)\citenamefont {Xu},
  \citenamefont {Zhong}, \citenamefont {Zhou}, \citenamefont {Han},
  \citenamefont {Jin}, \citenamefont {Gray}, \citenamefont {Jiang},\ and\
  \citenamefont {Zhang}}]{Jing2021}%
  \BibitemOpen
  \bibfield  {author} {\bibinfo {author} {\bibfnamefont {J.}~\bibnamefont
  {Xu}}, \bibinfo {author} {\bibfnamefont {C.}~\bibnamefont {Zhong}}, \bibinfo
  {author} {\bibfnamefont {X.}~\bibnamefont {Zhou}}, \bibinfo {author}
  {\bibfnamefont {X.}~\bibnamefont {Han}}, \bibinfo {author} {\bibfnamefont
  {D.}~\bibnamefont {Jin}}, \bibinfo {author} {\bibfnamefont {S.~K.}\
  \bibnamefont {Gray}}, \bibinfo {author} {\bibfnamefont {L.}~\bibnamefont
  {Jiang}}, \ and\ \bibinfo {author} {\bibfnamefont {X.}~\bibnamefont
  {Zhang}},\ }\href {\doibase 10.1103/PhysRevApplied.16.024009} {\bibfield
  {journal} {\bibinfo  {journal} {Phys. Rev. Applied}\ }\textbf {\bibinfo
  {volume} {16}},\ \bibinfo {pages} {024009} (\bibinfo {year}
  {2021})}\BibitemShut {NoStop}%
\bibitem [{\citenamefont {Shen}\ \emph {et~al.}(2021)\citenamefont {Shen},
  \citenamefont {Wang}, \citenamefont {Li}, \citenamefont {Zhu}, \citenamefont
  {Agarwal},\ and\ \citenamefont {You}}]{Shen2021}%
  \BibitemOpen
  \bibfield  {author} {\bibinfo {author} {\bibfnamefont {R.-C.}\ \bibnamefont
  {Shen}}, \bibinfo {author} {\bibfnamefont {Y.-P.}\ \bibnamefont {Wang}},
  \bibinfo {author} {\bibfnamefont {J.}~\bibnamefont {Li}}, \bibinfo {author}
  {\bibfnamefont {S.-Y.}\ \bibnamefont {Zhu}}, \bibinfo {author} {\bibfnamefont
  {G.~S.}\ \bibnamefont {Agarwal}}, \ and\ \bibinfo {author} {\bibfnamefont
  {J.~Q.}\ \bibnamefont {You}},\ }\href {\doibase
  10.1103/PhysRevLett.127.183202} {\bibfield  {journal} {\bibinfo  {journal}
  {Phys. Rev. Lett.}\ }\textbf {\bibinfo {volume} {127}},\ \bibinfo {pages}
  {183202} (\bibinfo {year} {2021})}\BibitemShut {NoStop}%
\bibitem [{\citenamefont {Kani}\ \emph {et~al.}(2022)\citenamefont {Kani},
  \citenamefont {Sarma},\ and\ \citenamefont {Twamley}}]{Kani2022}%
  \BibitemOpen
  \bibfield  {author} {\bibinfo {author} {\bibfnamefont {A.}~\bibnamefont
  {Kani}}, \bibinfo {author} {\bibfnamefont {B.}~\bibnamefont {Sarma}}, \ and\
  \bibinfo {author} {\bibfnamefont {J.}~\bibnamefont {Twamley}},\ }\href
  {\doibase 10.1103/physrevlett.128.013602} {\bibfield  {journal} {\bibinfo
  {journal} {Physical Review Letters}\ }\textbf {\bibinfo {volume} {128}},\
  \bibinfo {pages} {013602} (\bibinfo {year} {2022})}\BibitemShut {NoStop}%
\bibitem [{\citenamefont {Rameshti}\ \emph {et~al.}(2022)\citenamefont
  {Rameshti}, \citenamefont {Kusminskiy}, \citenamefont {Haigh}, \citenamefont
  {Usami}, \citenamefont {Lachance-Quirion}, \citenamefont {Nakamura},
  \citenamefont {Hu}, \citenamefont {Tang}, \citenamefont {Bauer},\ and\
  \citenamefont {Blanter}}]{Rameshti2022}%
  \BibitemOpen
  \bibfield  {author} {\bibinfo {author} {\bibfnamefont {B.~Z.}\ \bibnamefont
  {Rameshti}}, \bibinfo {author} {\bibfnamefont {S.~V.}\ \bibnamefont
  {Kusminskiy}}, \bibinfo {author} {\bibfnamefont {J.~A.}\ \bibnamefont
  {Haigh}}, \bibinfo {author} {\bibfnamefont {K.}~\bibnamefont {Usami}},
  \bibinfo {author} {\bibfnamefont {D.}~\bibnamefont {Lachance-Quirion}},
  \bibinfo {author} {\bibfnamefont {Y.}~\bibnamefont {Nakamura}}, \bibinfo
  {author} {\bibfnamefont {C.-M.}\ \bibnamefont {Hu}}, \bibinfo {author}
  {\bibfnamefont {H.~X.}\ \bibnamefont {Tang}}, \bibinfo {author}
  {\bibfnamefont {G.~E.}\ \bibnamefont {Bauer}}, \ and\ \bibinfo {author}
  {\bibfnamefont {Y.~M.}\ \bibnamefont {Blanter}},\ }\href {\doibase
  10.1016/j.physrep.2022.06.001} {\bibfield  {journal} {\bibinfo  {journal}
  {Physics Reports}\ }\textbf {\bibinfo {volume} {979}},\ \bibinfo {pages} {1}
  (\bibinfo {year} {2022})}\BibitemShut {NoStop}%
\bibitem [{\citenamefont {Huebl}\ \emph {et~al.}(2013)\citenamefont {Huebl},
  \citenamefont {Zollitsch}, \citenamefont {Lotze}, \citenamefont {Hocke},
  \citenamefont {Greifenstein}, \citenamefont {Marx}, \citenamefont {Gross},\
  and\ \citenamefont {Goennenwein}}]{Huebl2013}%
  \BibitemOpen
  \bibfield  {author} {\bibinfo {author} {\bibfnamefont {H.}~\bibnamefont
  {Huebl}}, \bibinfo {author} {\bibfnamefont {C.~W.}\ \bibnamefont
  {Zollitsch}}, \bibinfo {author} {\bibfnamefont {J.}~\bibnamefont {Lotze}},
  \bibinfo {author} {\bibfnamefont {F.}~\bibnamefont {Hocke}}, \bibinfo
  {author} {\bibfnamefont {M.}~\bibnamefont {Greifenstein}}, \bibinfo {author}
  {\bibfnamefont {A.}~\bibnamefont {Marx}}, \bibinfo {author} {\bibfnamefont
  {R.}~\bibnamefont {Gross}}, \ and\ \bibinfo {author} {\bibfnamefont
  {S.~T.~B.}\ \bibnamefont {Goennenwein}},\ }\href {\doibase
  10.1103/PhysRevLett.111.127003} {\bibfield  {journal} {\bibinfo  {journal}
  {Phys. Rev. Lett.}\ }\textbf {\bibinfo {volume} {111}},\ \bibinfo {pages}
  {127003} (\bibinfo {year} {2013})}\BibitemShut {NoStop}%
\bibitem [{\citenamefont {Tabuchi}\ \emph {et~al.}(2014)\citenamefont
  {Tabuchi}, \citenamefont {Ishino}, \citenamefont {Ishikawa}, \citenamefont
  {Yamazaki}, \citenamefont {Usami},\ and\ \citenamefont {Nakamura}}]{C1}%
  \BibitemOpen
  \bibfield  {author} {\bibinfo {author} {\bibfnamefont {Y.}~\bibnamefont
  {Tabuchi}}, \bibinfo {author} {\bibfnamefont {S.}~\bibnamefont {Ishino}},
  \bibinfo {author} {\bibfnamefont {T.}~\bibnamefont {Ishikawa}}, \bibinfo
  {author} {\bibfnamefont {R.}~\bibnamefont {Yamazaki}}, \bibinfo {author}
  {\bibfnamefont {K.}~\bibnamefont {Usami}}, \ and\ \bibinfo {author}
  {\bibfnamefont {Y.}~\bibnamefont {Nakamura}},\ }\href {\doibase
  10.1103/PhysRevLett.113.083603} {\bibfield  {journal} {\bibinfo  {journal}
  {Phys. Rev. Lett.}\ }\textbf {\bibinfo {volume} {113}},\ \bibinfo {pages}
  {083603} (\bibinfo {year} {2014})}\BibitemShut {NoStop}%
\bibitem [{\citenamefont {Zhang}\ \emph {et~al.}(2014)\citenamefont {Zhang},
  \citenamefont {Zou}, \citenamefont {Jiang},\ and\ \citenamefont {Tang}}]{C2}%
  \BibitemOpen
  \bibfield  {author} {\bibinfo {author} {\bibfnamefont {X.}~\bibnamefont
  {Zhang}}, \bibinfo {author} {\bibfnamefont {C.-L.}\ \bibnamefont {Zou}},
  \bibinfo {author} {\bibfnamefont {L.}~\bibnamefont {Jiang}}, \ and\ \bibinfo
  {author} {\bibfnamefont {H.~X.}\ \bibnamefont {Tang}},\ }\href {\doibase
  10.1103/physrevlett.113.156401} {\bibfield  {journal} {\bibinfo  {journal}
  {Physical Review Letters}\ }\textbf {\bibinfo {volume} {113}},\ \bibinfo
  {pages} {156401} (\bibinfo {year} {2014})}\BibitemShut {NoStop}%
\bibitem [{\citenamefont {Zhang}\ \emph {et~al.}(2015)\citenamefont {Zhang},
  \citenamefont {Wang}, \citenamefont {Li}, \citenamefont {Luo}, \citenamefont
  {Wu}, \citenamefont {Nori},\ and\ \citenamefont {You}}]{Zhang2015}%
  \BibitemOpen
  \bibfield  {author} {\bibinfo {author} {\bibfnamefont {D.}~\bibnamefont
  {Zhang}}, \bibinfo {author} {\bibfnamefont {X.-M.}\ \bibnamefont {Wang}},
  \bibinfo {author} {\bibfnamefont {T.-F.}\ \bibnamefont {Li}}, \bibinfo
  {author} {\bibfnamefont {X.-Q.}\ \bibnamefont {Luo}}, \bibinfo {author}
  {\bibfnamefont {W.}~\bibnamefont {Wu}}, \bibinfo {author} {\bibfnamefont
  {F.}~\bibnamefont {Nori}}, \ and\ \bibinfo {author} {\bibfnamefont
  {J.}~\bibnamefont {You}},\ }\href {\doibase 10.1038/npjqi.2015.14} {\bibfield
   {journal} {\bibinfo  {journal} {npj Quantum Information}\ }\textbf {\bibinfo
  {volume} {1}} (\bibinfo {year} {2015}),\ 10.1038/npjqi.2015.14}\BibitemShut
  {NoStop}%
\bibitem [{\citenamefont {Hou}\ and\ \citenamefont {Liu}(2019)}]{Hou2019}%
  \BibitemOpen
  \bibfield  {author} {\bibinfo {author} {\bibfnamefont {J.~T.}\ \bibnamefont
  {Hou}}\ and\ \bibinfo {author} {\bibfnamefont {L.}~\bibnamefont {Liu}},\
  }\href {\doibase 10.1103/PhysRevLett.123.107702} {\bibfield  {journal}
  {\bibinfo  {journal} {Phys. Rev. Lett.}\ }\textbf {\bibinfo {volume} {123}},\
  \bibinfo {pages} {107702} (\bibinfo {year} {2019})}\BibitemShut {NoStop}%
\bibitem [{\citenamefont {Li}\ \emph {et~al.}(2018)\citenamefont {Li},
  \citenamefont {Zhu},\ and\ \citenamefont {Agarwal}}]{D1}%
  \BibitemOpen
  \bibfield  {author} {\bibinfo {author} {\bibfnamefont {J.}~\bibnamefont
  {Li}}, \bibinfo {author} {\bibfnamefont {S.-Y.}\ \bibnamefont {Zhu}}, \ and\
  \bibinfo {author} {\bibfnamefont {G.~S.}\ \bibnamefont {Agarwal}},\ }\href
  {\doibase 10.1103/PhysRevLett.121.203601} {\bibfield  {journal} {\bibinfo
  {journal} {Phys. Rev. Lett.}\ }\textbf {\bibinfo {volume} {121}},\ \bibinfo
  {pages} {203601} (\bibinfo {year} {2018})}\BibitemShut {NoStop}%
\bibitem [{\citenamefont {Zhang}\ \emph {et~al.}(2019)\citenamefont {Zhang},
  \citenamefont {Scully},\ and\ \citenamefont {Agarwal}}]{D2}%
  \BibitemOpen
  \bibfield  {author} {\bibinfo {author} {\bibfnamefont {Z.}~\bibnamefont
  {Zhang}}, \bibinfo {author} {\bibfnamefont {M.~O.}\ \bibnamefont {Scully}}, \
  and\ \bibinfo {author} {\bibfnamefont {G.~S.}\ \bibnamefont {Agarwal}},\
  }\href {\doibase 10.1103/PhysRevResearch.1.023021} {\bibfield  {journal}
  {\bibinfo  {journal} {Phys. Rev. Research}\ }\textbf {\bibinfo {volume}
  {1}},\ \bibinfo {pages} {023021} (\bibinfo {year} {2019})}\BibitemShut
  {NoStop}%
\bibitem [{\citenamefont {Yuan}\ \emph {et~al.}(2020)\citenamefont {Yuan},
  \citenamefont {Yan}, \citenamefont {Zheng}, \citenamefont {He}, \citenamefont
  {Xia},\ and\ \citenamefont {Yung}}]{D6}%
  \BibitemOpen
  \bibfield  {author} {\bibinfo {author} {\bibfnamefont {H.~Y.}\ \bibnamefont
  {Yuan}}, \bibinfo {author} {\bibfnamefont {P.}~\bibnamefont {Yan}}, \bibinfo
  {author} {\bibfnamefont {S.}~\bibnamefont {Zheng}}, \bibinfo {author}
  {\bibfnamefont {Q.~Y.}\ \bibnamefont {He}}, \bibinfo {author} {\bibfnamefont
  {K.}~\bibnamefont {Xia}}, \ and\ \bibinfo {author} {\bibfnamefont {M.-H.}\
  \bibnamefont {Yung}},\ }\href {\doibase 10.1103/PhysRevLett.124.053602}
  {\bibfield  {journal} {\bibinfo  {journal} {Phys. Rev. Lett.}\ }\textbf
  {\bibinfo {volume} {124}},\ \bibinfo {pages} {053602} (\bibinfo {year}
  {2020})}\BibitemShut {NoStop}%
\bibitem [{\citenamefont {Nair}\ and\ \citenamefont {Agarwal}(2020)}]{D7}%
  \BibitemOpen
  \bibfield  {author} {\bibinfo {author} {\bibfnamefont {J.~M.~P.}\
  \bibnamefont {Nair}}\ and\ \bibinfo {author} {\bibfnamefont {G.~S.}\
  \bibnamefont {Agarwal}},\ }\href {\doibase 10.1063/5.0015195} {\bibfield
  {journal} {\bibinfo  {journal} {Applied Physics Letters}\ }\textbf {\bibinfo
  {volume} {117}},\ \bibinfo {pages} {084001} (\bibinfo {year}
  {2020})}\BibitemShut {NoStop}%
\bibitem [{\citenamefont {Yang}\ \emph {et~al.}(2021)\citenamefont {Yang},
  \citenamefont {Jin}, \citenamefont {Jin}, \citenamefont {Liu}, \citenamefont
  {Liu},\ and\ \citenamefont {Yang}}]{D8}%
  \BibitemOpen
  \bibfield  {author} {\bibinfo {author} {\bibfnamefont {Z.-B.}\ \bibnamefont
  {Yang}}, \bibinfo {author} {\bibfnamefont {H.}~\bibnamefont {Jin}}, \bibinfo
  {author} {\bibfnamefont {J.-W.}\ \bibnamefont {Jin}}, \bibinfo {author}
  {\bibfnamefont {J.-Y.}\ \bibnamefont {Liu}}, \bibinfo {author} {\bibfnamefont
  {H.-Y.}\ \bibnamefont {Liu}}, \ and\ \bibinfo {author} {\bibfnamefont
  {R.-C.}\ \bibnamefont {Yang}},\ }\href {\doibase
  10.1103/PhysRevResearch.3.023126} {\bibfield  {journal} {\bibinfo  {journal}
  {Phys. Rev. Research}\ }\textbf {\bibinfo {volume} {3}},\ \bibinfo {pages}
  {023126} (\bibinfo {year} {2021})}\BibitemShut {NoStop}%
\bibitem [{\citenamefont {Sun}\ \emph {et~al.}(2021)\citenamefont {Sun},
  \citenamefont {Zheng}, \citenamefont {Xiao}, \citenamefont {Gong},
  \citenamefont {He},\ and\ \citenamefont {Xia}}]{D9}%
  \BibitemOpen
  \bibfield  {author} {\bibinfo {author} {\bibfnamefont {F.-X.}\ \bibnamefont
  {Sun}}, \bibinfo {author} {\bibfnamefont {S.-S.}\ \bibnamefont {Zheng}},
  \bibinfo {author} {\bibfnamefont {Y.}~\bibnamefont {Xiao}}, \bibinfo {author}
  {\bibfnamefont {Q.}~\bibnamefont {Gong}}, \bibinfo {author} {\bibfnamefont
  {Q.}~\bibnamefont {He}}, \ and\ \bibinfo {author} {\bibfnamefont
  {K.}~\bibnamefont {Xia}},\ }\href {\doibase 10.1103/PhysRevLett.127.087203}
  {\bibfield  {journal} {\bibinfo  {journal} {Phys. Rev. Lett.}\ }\textbf
  {\bibinfo {volume} {127}},\ \bibinfo {pages} {087203} (\bibinfo {year}
  {2021})}\BibitemShut {NoStop}%
\bibitem [{\citenamefont {Azimi~Mousolou}\ \emph {et~al.}(2021)\citenamefont
  {Azimi~Mousolou}, \citenamefont {Liu}, \citenamefont {Bergman}, \citenamefont
  {Delin}, \citenamefont {Eriksson}, \citenamefont {Pereiro}, \citenamefont
  {Thonig},\ and\ \citenamefont {Sj\"oqvist}}]{D10}%
  \BibitemOpen
  \bibfield  {author} {\bibinfo {author} {\bibfnamefont {V.}~\bibnamefont
  {Azimi~Mousolou}}, \bibinfo {author} {\bibfnamefont {Y.}~\bibnamefont {Liu}},
  \bibinfo {author} {\bibfnamefont {A.}~\bibnamefont {Bergman}}, \bibinfo
  {author} {\bibfnamefont {A.}~\bibnamefont {Delin}}, \bibinfo {author}
  {\bibfnamefont {O.}~\bibnamefont {Eriksson}}, \bibinfo {author}
  {\bibfnamefont {M.}~\bibnamefont {Pereiro}}, \bibinfo {author} {\bibfnamefont
  {D.}~\bibnamefont {Thonig}}, \ and\ \bibinfo {author} {\bibfnamefont
  {E.}~\bibnamefont {Sj\"oqvist}},\ }\href {\doibase
  10.1103/PhysRevB.104.224302} {\bibfield  {journal} {\bibinfo  {journal}
  {Phys. Rev. B}\ }\textbf {\bibinfo {volume} {104}},\ \bibinfo {pages}
  {224302} (\bibinfo {year} {2021})}\BibitemShut {NoStop}%
\bibitem [{\citenamefont {Guan}\ \emph {et~al.}(2022)\citenamefont {Guan},
  \citenamefont {Wang},\ and\ \citenamefont {Yi}}]{Guan2022}%
  \BibitemOpen
  \bibfield  {author} {\bibinfo {author} {\bibfnamefont {S.-Y.}\ \bibnamefont
  {Guan}}, \bibinfo {author} {\bibfnamefont {H.-F.}\ \bibnamefont {Wang}}, \
  and\ \bibinfo {author} {\bibfnamefont {X.}~\bibnamefont {Yi}},\ }\href
  {\doibase 10.1038/s41534-022-00619-y} {\bibfield  {journal} {\bibinfo
  {journal} {npj Quantum Information}\ }\textbf {\bibinfo {volume} {8}}
  (\bibinfo {year} {2022}),\ 10.1038/s41534-022-00619-y}\BibitemShut {NoStop}%
\bibitem [{\citenamefont {Xu}\ \emph {et~al.}(2020)\citenamefont {Xu},
  \citenamefont {Zhong}, \citenamefont {Han}, \citenamefont {Jin},
  \citenamefont {Jiang},\ and\ \citenamefont {Zhang}}]{D11}%
  \BibitemOpen
  \bibfield  {author} {\bibinfo {author} {\bibfnamefont {J.}~\bibnamefont
  {Xu}}, \bibinfo {author} {\bibfnamefont {C.}~\bibnamefont {Zhong}}, \bibinfo
  {author} {\bibfnamefont {X.}~\bibnamefont {Han}}, \bibinfo {author}
  {\bibfnamefont {D.}~\bibnamefont {Jin}}, \bibinfo {author} {\bibfnamefont
  {L.}~\bibnamefont {Jiang}}, \ and\ \bibinfo {author} {\bibfnamefont
  {X.}~\bibnamefont {Zhang}},\ }\href {\doibase 10.1103/PhysRevLett.125.237201}
  {\bibfield  {journal} {\bibinfo  {journal} {Phys. Rev. Lett.}\ }\textbf
  {\bibinfo {volume} {125}},\ \bibinfo {pages} {237201} (\bibinfo {year}
  {2020})}\BibitemShut {NoStop}%
\bibitem [{\citenamefont {Bai}\ \emph {et~al.}(2017)\citenamefont {Bai},
  \citenamefont {Harder}, \citenamefont {Hyde}, \citenamefont {Zhang},
  \citenamefont {Hu}, \citenamefont {Chen},\ and\ \citenamefont {Xiao}}]{D12}%
  \BibitemOpen
  \bibfield  {author} {\bibinfo {author} {\bibfnamefont {L.}~\bibnamefont
  {Bai}}, \bibinfo {author} {\bibfnamefont {M.}~\bibnamefont {Harder}},
  \bibinfo {author} {\bibfnamefont {P.}~\bibnamefont {Hyde}}, \bibinfo {author}
  {\bibfnamefont {Z.}~\bibnamefont {Zhang}}, \bibinfo {author} {\bibfnamefont
  {C.-M.}\ \bibnamefont {Hu}}, \bibinfo {author} {\bibfnamefont {Y.~P.}\
  \bibnamefont {Chen}}, \ and\ \bibinfo {author} {\bibfnamefont {J.~Q.}\
  \bibnamefont {Xiao}},\ }\href {\doibase 10.1103/PhysRevLett.118.217201}
  {\bibfield  {journal} {\bibinfo  {journal} {Phys. Rev. Lett.}\ }\textbf
  {\bibinfo {volume} {118}},\ \bibinfo {pages} {217201} (\bibinfo {year}
  {2017})}\BibitemShut {NoStop}%
\bibitem [{\citenamefont {Nair}\ \emph {et~al.}(2020)\citenamefont {Nair},
  \citenamefont {Zhang}, \citenamefont {Scully},\ and\ \citenamefont
  {Agarwal}}]{D13}%
  \BibitemOpen
  \bibfield  {author} {\bibinfo {author} {\bibfnamefont {J.~M.~P.}\
  \bibnamefont {Nair}}, \bibinfo {author} {\bibfnamefont {Z.}~\bibnamefont
  {Zhang}}, \bibinfo {author} {\bibfnamefont {M.~O.}\ \bibnamefont {Scully}}, \
  and\ \bibinfo {author} {\bibfnamefont {G.~S.}\ \bibnamefont {Agarwal}},\
  }\href {\doibase 10.1103/PhysRevB.102.104415} {\bibfield  {journal} {\bibinfo
   {journal} {Phys. Rev. B}\ }\textbf {\bibinfo {volume} {102}},\ \bibinfo
  {pages} {104415} (\bibinfo {year} {2020})}\BibitemShut {NoStop}%
\bibitem [{\citenamefont {Kong}\ \emph {et~al.}(2019)\citenamefont {Kong},
  \citenamefont {Xiong},\ and\ \citenamefont {Wu}}]{D14}%
  \BibitemOpen
  \bibfield  {author} {\bibinfo {author} {\bibfnamefont {C.}~\bibnamefont
  {Kong}}, \bibinfo {author} {\bibfnamefont {H.}~\bibnamefont {Xiong}}, \ and\
  \bibinfo {author} {\bibfnamefont {Y.}~\bibnamefont {Wu}},\ }\href {\doibase
  10.1103/physrevapplied.12.034001} {\bibfield  {journal} {\bibinfo  {journal}
  {Physical Review Applied}\ }\textbf {\bibinfo {volume} {12}},\ \bibinfo
  {pages} {034001} (\bibinfo {year} {2019})}\BibitemShut {NoStop}%
\bibitem [{\citenamefont {Wang}\ \emph {et~al.}(2019)\citenamefont {Wang},
  \citenamefont {Rao}, \citenamefont {Yang}, \citenamefont {Xu}, \citenamefont
  {Gui}, \citenamefont {Yao}, \citenamefont {You},\ and\ \citenamefont
  {Hu}}]{D15}%
  \BibitemOpen
  \bibfield  {author} {\bibinfo {author} {\bibfnamefont {Y.-P.}\ \bibnamefont
  {Wang}}, \bibinfo {author} {\bibfnamefont {J.}~\bibnamefont {Rao}}, \bibinfo
  {author} {\bibfnamefont {Y.}~\bibnamefont {Yang}}, \bibinfo {author}
  {\bibfnamefont {P.-C.}\ \bibnamefont {Xu}}, \bibinfo {author} {\bibfnamefont
  {Y.}~\bibnamefont {Gui}}, \bibinfo {author} {\bibfnamefont {B.}~\bibnamefont
  {Yao}}, \bibinfo {author} {\bibfnamefont {J.}~\bibnamefont {You}}, \ and\
  \bibinfo {author} {\bibfnamefont {C.-M.}\ \bibnamefont {Hu}},\ }\href
  {\doibase 10.1103/physrevlett.123.127202} {\bibfield  {journal} {\bibinfo
  {journal} {Physical Review Letters}\ }\textbf {\bibinfo {volume} {123}},\
  \bibinfo {pages} {127202} (\bibinfo {year} {2019})}\BibitemShut {NoStop}%
\bibitem [{\citenamefont {Zhu}\ \emph {et~al.}(2020)\citenamefont {Zhu},
  \citenamefont {Han}, \citenamefont {Zou}, \citenamefont {Xu},\ and\
  \citenamefont {Tang}}]{D16}%
  \BibitemOpen
  \bibfield  {author} {\bibinfo {author} {\bibfnamefont {N.}~\bibnamefont
  {Zhu}}, \bibinfo {author} {\bibfnamefont {X.}~\bibnamefont {Han}}, \bibinfo
  {author} {\bibfnamefont {C.-L.}\ \bibnamefont {Zou}}, \bibinfo {author}
  {\bibfnamefont {M.}~\bibnamefont {Xu}}, \ and\ \bibinfo {author}
  {\bibfnamefont {H.~X.}\ \bibnamefont {Tang}},\ }\href {\doibase
  10.1103/physreva.101.043842} {\bibfield  {journal} {\bibinfo  {journal}
  {Physical Review A}\ }\textbf {\bibinfo {volume} {101}},\ \bibinfo {pages}
  {043842} (\bibinfo {year} {2020})}\BibitemShut {NoStop}%
\bibitem [{\citenamefont {Zhang}\ \emph {et~al.}(2020)\citenamefont {Zhang},
  \citenamefont {Galda}, \citenamefont {Han}, \citenamefont {Jin},\ and\
  \citenamefont {Vinokur}}]{D17}%
  \BibitemOpen
  \bibfield  {author} {\bibinfo {author} {\bibfnamefont {X.}~\bibnamefont
  {Zhang}}, \bibinfo {author} {\bibfnamefont {A.}~\bibnamefont {Galda}},
  \bibinfo {author} {\bibfnamefont {X.}~\bibnamefont {Han}}, \bibinfo {author}
  {\bibfnamefont {D.}~\bibnamefont {Jin}}, \ and\ \bibinfo {author}
  {\bibfnamefont {V.~M.}\ \bibnamefont {Vinokur}},\ }\href {\doibase
  10.1103/physrevapplied.13.044039} {\bibfield  {journal} {\bibinfo  {journal}
  {Physical Review Applied}\ }\textbf {\bibinfo {volume} {13}},\ \bibinfo
  {pages} {044039} (\bibinfo {year} {2020})}\BibitemShut {NoStop}%
\bibitem [{\citenamefont {Yu}\ \emph {et~al.}(2020)\citenamefont {Yu},
  \citenamefont {Zhang}, \citenamefont {Sharma}, \citenamefont {Zhang},
  \citenamefont {Blanter},\ and\ \citenamefont {Bauer}}]{D18}%
  \BibitemOpen
  \bibfield  {author} {\bibinfo {author} {\bibfnamefont {T.}~\bibnamefont
  {Yu}}, \bibinfo {author} {\bibfnamefont {Y.-X.}\ \bibnamefont {Zhang}},
  \bibinfo {author} {\bibfnamefont {S.}~\bibnamefont {Sharma}}, \bibinfo
  {author} {\bibfnamefont {X.}~\bibnamefont {Zhang}}, \bibinfo {author}
  {\bibfnamefont {Y.~M.}\ \bibnamefont {Blanter}}, \ and\ \bibinfo {author}
  {\bibfnamefont {G.~E.}\ \bibnamefont {Bauer}},\ }\href {\doibase
  10.1103/physrevlett.124.107202} {\bibfield  {journal} {\bibinfo  {journal}
  {Physical Review Letters}\ }\textbf {\bibinfo {volume} {124}},\ \bibinfo
  {pages} {107202} (\bibinfo {year} {2020})}\BibitemShut {NoStop}%
\bibitem [{\citenamefont {Ren}\ \emph {et~al.}(2022{\natexlab{a}})\citenamefont
  {Ren}, \citenamefont {Ma}, \citenamefont {Xie}, \citenamefont {Li},
  \citenamefont {Cao},\ and\ \citenamefont {Li}}]{RenR}%
  \BibitemOpen
  \bibfield  {author} {\bibinfo {author} {\bibfnamefont {Y.-l.}\ \bibnamefont
  {Ren}}, \bibinfo {author} {\bibfnamefont {S.-l.}\ \bibnamefont {Ma}},
  \bibinfo {author} {\bibfnamefont {J.-k.}\ \bibnamefont {Xie}}, \bibinfo
  {author} {\bibfnamefont {X.-k.}\ \bibnamefont {Li}}, \bibinfo {author}
  {\bibfnamefont {M.-t.}\ \bibnamefont {Cao}}, \ and\ \bibinfo {author}
  {\bibfnamefont {F.-l.}\ \bibnamefont {Li}},\ }\href {\doibase
  10.1103/PhysRevA.105.013711} {\bibfield  {journal} {\bibinfo  {journal}
  {Phys. Rev. A}\ }\textbf {\bibinfo {volume} {105}},\ \bibinfo {pages}
  {013711} (\bibinfo {year} {2022}{\natexlab{a}})}\BibitemShut {NoStop}%
\bibitem [{\citenamefont {Zhang}\ \emph {et~al.}(2017)\citenamefont {Zhang},
  \citenamefont {Luo}, \citenamefont {Wang}, \citenamefont {Li},\ and\
  \citenamefont {You}}]{Zhang2017}%
  \BibitemOpen
  \bibfield  {author} {\bibinfo {author} {\bibfnamefont {D.}~\bibnamefont
  {Zhang}}, \bibinfo {author} {\bibfnamefont {X.-Q.}\ \bibnamefont {Luo}},
  \bibinfo {author} {\bibfnamefont {Y.-P.}\ \bibnamefont {Wang}}, \bibinfo
  {author} {\bibfnamefont {T.-F.}\ \bibnamefont {Li}}, \ and\ \bibinfo {author}
  {\bibfnamefont {J.~Q.}\ \bibnamefont {You}},\ }\href {\doibase
  10.1038/s41467-017-01634-w} {\bibfield  {journal} {\bibinfo  {journal}
  {Nature Communications}\ }\textbf {\bibinfo {volume} {8}} (\bibinfo {year}
  {2017}),\ 10.1038/s41467-017-01634-w}\BibitemShut {NoStop}%
\bibitem [{\citenamefont {Harder}\ \emph {et~al.}(2018)\citenamefont {Harder},
  \citenamefont {Yang}, \citenamefont {Yao}, \citenamefont {Yu}, \citenamefont
  {Rao}, \citenamefont {Gui}, \citenamefont {Stamps},\ and\ \citenamefont
  {Hu}}]{D20}%
  \BibitemOpen
  \bibfield  {author} {\bibinfo {author} {\bibfnamefont {M.}~\bibnamefont
  {Harder}}, \bibinfo {author} {\bibfnamefont {Y.}~\bibnamefont {Yang}},
  \bibinfo {author} {\bibfnamefont {B.}~\bibnamefont {Yao}}, \bibinfo {author}
  {\bibfnamefont {C.}~\bibnamefont {Yu}}, \bibinfo {author} {\bibfnamefont
  {J.}~\bibnamefont {Rao}}, \bibinfo {author} {\bibfnamefont {Y.}~\bibnamefont
  {Gui}}, \bibinfo {author} {\bibfnamefont {R.}~\bibnamefont {Stamps}}, \ and\
  \bibinfo {author} {\bibfnamefont {C.-M.}\ \bibnamefont {Hu}},\ }\href
  {\doibase 10.1103/physrevlett.121.137203} {\bibfield  {journal} {\bibinfo
  {journal} {Physical Review Letters}\ }\textbf {\bibinfo {volume} {121}},\
  \bibinfo {pages} {137203} (\bibinfo {year} {2018})}\BibitemShut {NoStop}%
\bibitem [{\citenamefont {Liu}\ \emph {et~al.}(2019{\natexlab{a}})\citenamefont
  {Liu}, \citenamefont {Sun}, \citenamefont {Zhang}, \citenamefont {Groesbeck},
  \citenamefont {Mclaughlin},\ and\ \citenamefont {Vardeny}}]{Liu2019}%
  \BibitemOpen
  \bibfield  {author} {\bibinfo {author} {\bibfnamefont {H.}~\bibnamefont
  {Liu}}, \bibinfo {author} {\bibfnamefont {D.}~\bibnamefont {Sun}}, \bibinfo
  {author} {\bibfnamefont {C.}~\bibnamefont {Zhang}}, \bibinfo {author}
  {\bibfnamefont {M.}~\bibnamefont {Groesbeck}}, \bibinfo {author}
  {\bibfnamefont {R.}~\bibnamefont {Mclaughlin}}, \ and\ \bibinfo {author}
  {\bibfnamefont {Z.~V.}\ \bibnamefont {Vardeny}},\ }\href {\doibase
  10.1126/sciadv.aax9144} {\bibfield  {journal} {\bibinfo  {journal} {Science
  Advances}\ }\textbf {\bibinfo {volume} {5}} (\bibinfo {year}
  {2019}{\natexlab{a}}),\ 10.1126/sciadv.aax9144}\BibitemShut {NoStop}%
\bibitem [{\citenamefont {Xu}\ \emph {et~al.}(2019)\citenamefont {Xu},
  \citenamefont {Rao}, \citenamefont {Gui}, \citenamefont {Jin},\ and\
  \citenamefont {Hu}}]{D21}%
  \BibitemOpen
  \bibfield  {author} {\bibinfo {author} {\bibfnamefont {P.-C.}\ \bibnamefont
  {Xu}}, \bibinfo {author} {\bibfnamefont {J.~W.}\ \bibnamefont {Rao}},
  \bibinfo {author} {\bibfnamefont {Y.~S.}\ \bibnamefont {Gui}}, \bibinfo
  {author} {\bibfnamefont {X.}~\bibnamefont {Jin}}, \ and\ \bibinfo {author}
  {\bibfnamefont {C.-M.}\ \bibnamefont {Hu}},\ }\href {\doibase
  10.1103/physrevb.100.094415} {\bibfield  {journal} {\bibinfo  {journal}
  {Physical Review B}\ }\textbf {\bibinfo {volume} {100}},\ \bibinfo {pages}
  {094415} (\bibinfo {year} {2019})}\BibitemShut {NoStop}%
\bibitem [{\citenamefont {Yang}\ \emph {et~al.}(2020)\citenamefont {Yang},
  \citenamefont {Wang}, \citenamefont {Rao}, \citenamefont {Gui}, \citenamefont
  {Yao}, \citenamefont {Lu},\ and\ \citenamefont {Hu}}]{D22}%
  \BibitemOpen
  \bibfield  {author} {\bibinfo {author} {\bibfnamefont {Y.}~\bibnamefont
  {Yang}}, \bibinfo {author} {\bibfnamefont {Y.-P.}\ \bibnamefont {Wang}},
  \bibinfo {author} {\bibfnamefont {J.}~\bibnamefont {Rao}}, \bibinfo {author}
  {\bibfnamefont {Y.}~\bibnamefont {Gui}}, \bibinfo {author} {\bibfnamefont
  {B.}~\bibnamefont {Yao}}, \bibinfo {author} {\bibfnamefont {W.}~\bibnamefont
  {Lu}}, \ and\ \bibinfo {author} {\bibfnamefont {C.-M.}\ \bibnamefont {Hu}},\
  }\href {\doibase 10.1103/physrevlett.125.147202} {\bibfield  {journal}
  {\bibinfo  {journal} {Physical Review Letters}\ }\textbf {\bibinfo {volume}
  {125}},\ \bibinfo {pages} {147202} (\bibinfo {year} {2020})}\BibitemShut
  {NoStop}%
\bibitem [{\citenamefont {Lu}\ \emph {et~al.}(2021)\citenamefont {Lu},
  \citenamefont {Zhang}, \citenamefont {Zhang},\ and\ \citenamefont
  {Jing}}]{D24}%
  \BibitemOpen
  \bibfield  {author} {\bibinfo {author} {\bibfnamefont {T.-X.}\ \bibnamefont
  {Lu}}, \bibinfo {author} {\bibfnamefont {H.}~\bibnamefont {Zhang}}, \bibinfo
  {author} {\bibfnamefont {Q.}~\bibnamefont {Zhang}}, \ and\ \bibinfo {author}
  {\bibfnamefont {H.}~\bibnamefont {Jing}},\ }\href {\doibase
  10.1103/PhysRevA.103.063708} {\bibfield  {journal} {\bibinfo  {journal}
  {Phys. Rev. A}\ }\textbf {\bibinfo {volume} {103}},\ \bibinfo {pages}
  {063708} (\bibinfo {year} {2021})}\BibitemShut {NoStop}%
\bibitem [{\citenamefont {Flower}\ \emph {et~al.}(2019)\citenamefont {Flower},
  \citenamefont {Bourhill}, \citenamefont {Goryachev},\ and\ \citenamefont
  {Tobar}}]{D25}%
  \BibitemOpen
  \bibfield  {author} {\bibinfo {author} {\bibfnamefont {G.}~\bibnamefont
  {Flower}}, \bibinfo {author} {\bibfnamefont {J.}~\bibnamefont {Bourhill}},
  \bibinfo {author} {\bibfnamefont {M.}~\bibnamefont {Goryachev}}, \ and\
  \bibinfo {author} {\bibfnamefont {M.~E.}\ \bibnamefont {Tobar}},\ }\href
  {\doibase 10.1016/j.dark.2019.100306} {\bibfield  {journal} {\bibinfo
  {journal} {Physics of the Dark Universe}\ }\textbf {\bibinfo {volume} {25}},\
  \bibinfo {pages} {100306} (\bibinfo {year} {2019})}\BibitemShut {NoStop}%
\bibitem [{\citenamefont {Crescini}\ \emph
  {et~al.}(2020{\natexlab{a}})\citenamefont {Crescini}, \citenamefont
  {Alesini}, \citenamefont {Braggio}, \citenamefont {Carugno}, \citenamefont
  {D'Agostino}, \citenamefont {Gioacchino}, \citenamefont {Falferi},
  \citenamefont {Gambardella}, \citenamefont {Gatti}, \citenamefont {Iannone},
  \citenamefont {Ligi}, \citenamefont {Lombardi}, \citenamefont {Ortolan},
  \citenamefont {Pengo}, \citenamefont {Ruoso},\ and\ \citenamefont
  {and}}]{D26}%
  \BibitemOpen
  \bibfield  {author} {\bibinfo {author} {\bibfnamefont {N.}~\bibnamefont
  {Crescini}}, \bibinfo {author} {\bibfnamefont {D.}~\bibnamefont {Alesini}},
  \bibinfo {author} {\bibfnamefont {C.}~\bibnamefont {Braggio}}, \bibinfo
  {author} {\bibfnamefont {G.}~\bibnamefont {Carugno}}, \bibinfo {author}
  {\bibfnamefont {D.}~\bibnamefont {D'Agostino}}, \bibinfo {author}
  {\bibfnamefont {D.~D.}\ \bibnamefont {Gioacchino}}, \bibinfo {author}
  {\bibfnamefont {P.}~\bibnamefont {Falferi}}, \bibinfo {author} {\bibfnamefont
  {U.}~\bibnamefont {Gambardella}}, \bibinfo {author} {\bibfnamefont
  {C.}~\bibnamefont {Gatti}}, \bibinfo {author} {\bibfnamefont
  {G.}~\bibnamefont {Iannone}}, \bibinfo {author} {\bibfnamefont
  {C.}~\bibnamefont {Ligi}}, \bibinfo {author} {\bibfnamefont {A.}~\bibnamefont
  {Lombardi}}, \bibinfo {author} {\bibfnamefont {A.}~\bibnamefont {Ortolan}},
  \bibinfo {author} {\bibfnamefont {R.}~\bibnamefont {Pengo}}, \bibinfo
  {author} {\bibfnamefont {G.}~\bibnamefont {Ruoso}}, \ and\ \bibinfo {author}
  {\bibfnamefont {L.~T.}\ \bibnamefont {and}},\ }\href {\doibase
  10.1103/physrevlett.124.171801} {\bibfield  {journal} {\bibinfo  {journal}
  {Physical Review Letters}\ }\textbf {\bibinfo {volume} {124}},\ \bibinfo
  {pages} {171801} (\bibinfo {year} {2020}{\natexlab{a}})}\BibitemShut
  {NoStop}%
\bibitem [{\citenamefont {Crescini}\ \emph
  {et~al.}(2020{\natexlab{b}})\citenamefont {Crescini}, \citenamefont
  {Braggio}, \citenamefont {Carugno}, \citenamefont {Vora}, \citenamefont
  {Ortolan},\ and\ \citenamefont {Ruoso}}]{D27}%
  \BibitemOpen
  \bibfield  {author} {\bibinfo {author} {\bibfnamefont {N.}~\bibnamefont
  {Crescini}}, \bibinfo {author} {\bibfnamefont {C.}~\bibnamefont {Braggio}},
  \bibinfo {author} {\bibfnamefont {G.}~\bibnamefont {Carugno}}, \bibinfo
  {author} {\bibfnamefont {R.~D.}\ \bibnamefont {Vora}}, \bibinfo {author}
  {\bibfnamefont {A.}~\bibnamefont {Ortolan}}, \ and\ \bibinfo {author}
  {\bibfnamefont {G.}~\bibnamefont {Ruoso}},\ }\href {\doibase
  10.1038/s42005-020-00435-w} {\bibfield  {journal} {\bibinfo  {journal}
  {Communications Physics}\ }\textbf {\bibinfo {volume} {3}} (\bibinfo {year}
  {2020}{\natexlab{b}}),\ 10.1038/s42005-020-00435-w}\BibitemShut {NoStop}%
\bibitem [{\citenamefont {Tabuchi}\ \emph {et~al.}(2015)\citenamefont
  {Tabuchi}, \citenamefont {Ishino}, \citenamefont {Noguchi}, \citenamefont
  {Ishikawa}, \citenamefont {Yamazaki}, \citenamefont {Usami},\ and\
  \citenamefont {Nakamura}}]{E2}%
  \BibitemOpen
  \bibfield  {author} {\bibinfo {author} {\bibfnamefont {Y.}~\bibnamefont
  {Tabuchi}}, \bibinfo {author} {\bibfnamefont {S.}~\bibnamefont {Ishino}},
  \bibinfo {author} {\bibfnamefont {A.}~\bibnamefont {Noguchi}}, \bibinfo
  {author} {\bibfnamefont {T.}~\bibnamefont {Ishikawa}}, \bibinfo {author}
  {\bibfnamefont {R.}~\bibnamefont {Yamazaki}}, \bibinfo {author}
  {\bibfnamefont {K.}~\bibnamefont {Usami}}, \ and\ \bibinfo {author}
  {\bibfnamefont {Y.}~\bibnamefont {Nakamura}},\ }\href {\doibase
  10.1126/science.aaa3693} {\bibfield  {journal} {\bibinfo  {journal}
  {Science}\ }\textbf {\bibinfo {volume} {349}},\ \bibinfo {pages} {405}
  (\bibinfo {year} {2015})}\BibitemShut {NoStop}%
\bibitem [{\citenamefont {Yuan}\ \emph {et~al.}(2022)\citenamefont {Yuan},
  \citenamefont {Cao}, \citenamefont {Kamra}, \citenamefont {Duine},\ and\
  \citenamefont {Yan}}]{B4}%
  \BibitemOpen
  \bibfield  {author} {\bibinfo {author} {\bibfnamefont {H.}~\bibnamefont
  {Yuan}}, \bibinfo {author} {\bibfnamefont {Y.}~\bibnamefont {Cao}}, \bibinfo
  {author} {\bibfnamefont {A.}~\bibnamefont {Kamra}}, \bibinfo {author}
  {\bibfnamefont {R.~A.}\ \bibnamefont {Duine}}, \ and\ \bibinfo {author}
  {\bibfnamefont {P.}~\bibnamefont {Yan}},\ }\href {\doibase
  10.1016/j.physrep.2022.03.002} {\bibfield  {journal} {\bibinfo  {journal}
  {Physics Reports}\ }\textbf {\bibinfo {volume} {965}},\ \bibinfo {pages} {1}
  (\bibinfo {year} {2022})}\BibitemShut {NoStop}%
\bibitem [{\citenamefont {Liu}\ \emph {et~al.}(2019{\natexlab{b}})\citenamefont
  {Liu}, \citenamefont {Xiong},\ and\ \citenamefont {Wu}}]{E8}%
  \BibitemOpen
  \bibfield  {author} {\bibinfo {author} {\bibfnamefont {Z.-X.}\ \bibnamefont
  {Liu}}, \bibinfo {author} {\bibfnamefont {H.}~\bibnamefont {Xiong}}, \ and\
  \bibinfo {author} {\bibfnamefont {Y.}~\bibnamefont {Wu}},\ }\href {\doibase
  10.1103/PhysRevB.100.134421} {\bibfield  {journal} {\bibinfo  {journal}
  {Phys. Rev. B}\ }\textbf {\bibinfo {volume} {100}},\ \bibinfo {pages}
  {134421} (\bibinfo {year} {2019}{\natexlab{b}})}\BibitemShut {NoStop}%
\bibitem [{\citenamefont {Xie}\ \emph {et~al.}(2020)\citenamefont {Xie},
  \citenamefont {Ma},\ and\ \citenamefont {Li}}]{E9}%
  \BibitemOpen
  \bibfield  {author} {\bibinfo {author} {\bibfnamefont {J.-k.}\ \bibnamefont
  {Xie}}, \bibinfo {author} {\bibfnamefont {S.-l.}\ \bibnamefont {Ma}}, \ and\
  \bibinfo {author} {\bibfnamefont {F.-l.}\ \bibnamefont {Li}},\ }\href
  {\doibase 10.1103/PhysRevA.101.042331} {\bibfield  {journal} {\bibinfo
  {journal} {Phys. Rev. A}\ }\textbf {\bibinfo {volume} {101}},\ \bibinfo
  {pages} {042331} (\bibinfo {year} {2020})}\BibitemShut {NoStop}%
\bibitem [{\citenamefont {Li}\ \emph {et~al.}(2021)\citenamefont {Li},
  \citenamefont {Wang}, \citenamefont {Wu}, \citenamefont {Yang},\ and\
  \citenamefont {Chen}}]{E10}%
  \BibitemOpen
  \bibfield  {author} {\bibinfo {author} {\bibfnamefont {X.}~\bibnamefont
  {Li}}, \bibinfo {author} {\bibfnamefont {X.}~\bibnamefont {Wang}}, \bibinfo
  {author} {\bibfnamefont {Z.}~\bibnamefont {Wu}}, \bibinfo {author}
  {\bibfnamefont {W.-X.}\ \bibnamefont {Yang}}, \ and\ \bibinfo {author}
  {\bibfnamefont {A.}~\bibnamefont {Chen}},\ }\href {\doibase
  10.1103/physrevb.104.224434} {\bibfield  {journal} {\bibinfo  {journal}
  {Physical Review B}\ }\textbf {\bibinfo {volume} {104}},\ \bibinfo {pages}
  {224434} (\bibinfo {year} {2021})}\BibitemShut {NoStop}%
\bibitem [{\citenamefont {Kong}\ \emph {et~al.}(2022)\citenamefont {Kong},
  \citenamefont {Xu}, \citenamefont {Tian}, \citenamefont {Wang},\ and\
  \citenamefont {Hu}}]{E11}%
  \BibitemOpen
  \bibfield  {author} {\bibinfo {author} {\bibfnamefont {D.}~\bibnamefont
  {Kong}}, \bibinfo {author} {\bibfnamefont {J.}~\bibnamefont {Xu}}, \bibinfo
  {author} {\bibfnamefont {Y.}~\bibnamefont {Tian}}, \bibinfo {author}
  {\bibfnamefont {F.}~\bibnamefont {Wang}}, \ and\ \bibinfo {author}
  {\bibfnamefont {X.}~\bibnamefont {Hu}},\ }\href {\doibase
  10.1103/physrevresearch.4.013084} {\bibfield  {journal} {\bibinfo  {journal}
  {Physical Review Research}\ }\textbf {\bibinfo {volume} {4}},\ \bibinfo
  {pages} {013084} (\bibinfo {year} {2022})}\BibitemShut {NoStop}%
\bibitem [{\citenamefont {fan Qi}\ and\ \citenamefont {Jing}(2022)}]{E12}%
  \BibitemOpen
  \bibfield  {author} {\bibinfo {author} {\bibfnamefont {S.}~\bibnamefont {fan
  Qi}}\ and\ \bibinfo {author} {\bibfnamefont {J.}~\bibnamefont {Jing}},\
  }\href {\doibase 10.1103/physreva.105.022624} {\bibfield  {journal} {\bibinfo
   {journal} {Physical Review A}\ }\textbf {\bibinfo {volume} {105}},\ \bibinfo
  {pages} {022624} (\bibinfo {year} {2022})}\BibitemShut {NoStop}%
\bibitem [{\citenamefont {Ren}\ \emph {et~al.}(2022{\natexlab{b}})\citenamefont
  {Ren}, \citenamefont {Xie}, \citenamefont {Li}, \citenamefont {Ma},\ and\
  \citenamefont {Li}}]{E14}%
  \BibitemOpen
  \bibfield  {author} {\bibinfo {author} {\bibfnamefont {Y.-l.}\ \bibnamefont
  {Ren}}, \bibinfo {author} {\bibfnamefont {J.-k.}\ \bibnamefont {Xie}},
  \bibinfo {author} {\bibfnamefont {X.-k.}\ \bibnamefont {Li}}, \bibinfo
  {author} {\bibfnamefont {S.-l.}\ \bibnamefont {Ma}}, \ and\ \bibinfo {author}
  {\bibfnamefont {F.-l.}\ \bibnamefont {Li}},\ }\href {\doibase
  10.1103/PhysRevB.105.094422} {\bibfield  {journal} {\bibinfo  {journal}
  {Phys. Rev. B}\ }\textbf {\bibinfo {volume} {105}},\ \bibinfo {pages}
  {094422} (\bibinfo {year} {2022}{\natexlab{b}})}\BibitemShut {NoStop}%
\bibitem [{\citenamefont {Sharma}\ \emph {et~al.}(2022)\citenamefont {Sharma},
  \citenamefont {Bittencourt},\ and\ \citenamefont {Kusminskiy}}]{Sharma20222}%
  \BibitemOpen
  \bibfield  {author} {\bibinfo {author} {\bibfnamefont {S.}~\bibnamefont
  {Sharma}}, \bibinfo {author} {\bibfnamefont {V.~S.~V.}\ \bibnamefont
  {Bittencourt}}, \ and\ \bibinfo {author} {\bibfnamefont {S.~V.}\ \bibnamefont
  {Kusminskiy}},\ }\href {\doibase 10.1088/2515-7639/ac81f0} {\bibfield
  {journal} {\bibinfo  {journal} {Journal of Physics: Materials}\ }\textbf
  {\bibinfo {volume} {5}},\ \bibinfo {pages} {034006} (\bibinfo {year}
  {2022})}\BibitemShut {NoStop}%
\bibitem [{\citenamefont {Kounalakis}\ \emph {et~al.}(2022)\citenamefont
  {Kounalakis}, \citenamefont {Bauer},\ and\ \citenamefont
  {Blanter}}]{Kounalakis20222}%
  \BibitemOpen
  \bibfield  {author} {\bibinfo {author} {\bibfnamefont {M.}~\bibnamefont
  {Kounalakis}}, \bibinfo {author} {\bibfnamefont {G.~E.~W.}\ \bibnamefont
  {Bauer}}, \ and\ \bibinfo {author} {\bibfnamefont {Y.~M.}\ \bibnamefont
  {Blanter}},\ }\href {\doibase 10.1103/PhysRevLett.129.037205} {\bibfield
  {journal} {\bibinfo  {journal} {Phys. Rev. Lett.}\ }\textbf {\bibinfo
  {volume} {129}},\ \bibinfo {pages} {037205} (\bibinfo {year}
  {2022})}\BibitemShut {NoStop}%
\bibitem [{\citenamefont {Lachance-Quirion}\ \emph {et~al.}(2020)\citenamefont
  {Lachance-Quirion}, \citenamefont {Wolski}, \citenamefont {Tabuchi},
  \citenamefont {Kono}, \citenamefont {Usami},\ and\ \citenamefont
  {Nakamura}}]{E6}%
  \BibitemOpen
  \bibfield  {author} {\bibinfo {author} {\bibfnamefont {D.}~\bibnamefont
  {Lachance-Quirion}}, \bibinfo {author} {\bibfnamefont {S.~P.}\ \bibnamefont
  {Wolski}}, \bibinfo {author} {\bibfnamefont {Y.}~\bibnamefont {Tabuchi}},
  \bibinfo {author} {\bibfnamefont {S.}~\bibnamefont {Kono}}, \bibinfo {author}
  {\bibfnamefont {K.}~\bibnamefont {Usami}}, \ and\ \bibinfo {author}
  {\bibfnamefont {Y.}~\bibnamefont {Nakamura}},\ }\href {\doibase
  10.1126/science.aaz9236} {\bibfield  {journal} {\bibinfo  {journal}
  {Science}\ }\textbf {\bibinfo {volume} {367}},\ \bibinfo {pages} {425}
  (\bibinfo {year} {2020})}\BibitemShut {NoStop}%
\bibitem [{\citenamefont {Zhou}\ \emph {et~al.}(2008)\citenamefont {Zhou},
  \citenamefont {Gong}, \citenamefont {Liu}, \citenamefont {Sun},\ and\
  \citenamefont {Nori}}]{Zhou2008}%
  \BibitemOpen
  \bibfield  {author} {\bibinfo {author} {\bibfnamefont {L.}~\bibnamefont
  {Zhou}}, \bibinfo {author} {\bibfnamefont {Z.~R.}\ \bibnamefont {Gong}},
  \bibinfo {author} {\bibfnamefont {Y.-x.}\ \bibnamefont {Liu}}, \bibinfo
  {author} {\bibfnamefont {C.~P.}\ \bibnamefont {Sun}}, \ and\ \bibinfo
  {author} {\bibfnamefont {F.}~\bibnamefont {Nori}},\ }\href {\doibase
  10.1103/PhysRevLett.101.100501} {\bibfield  {journal} {\bibinfo  {journal}
  {Phys. Rev. Lett.}\ }\textbf {\bibinfo {volume} {101}},\ \bibinfo {pages}
  {100501} (\bibinfo {year} {2008})}\BibitemShut {NoStop}%
\bibitem [{\citenamefont {Fang}\ \emph {et~al.}(2012)\citenamefont {Fang},
  \citenamefont {Yu},\ and\ \citenamefont {Fan}}]{Fang2012}%
  \BibitemOpen
  \bibfield  {author} {\bibinfo {author} {\bibfnamefont {K.}~\bibnamefont
  {Fang}}, \bibinfo {author} {\bibfnamefont {Z.}~\bibnamefont {Yu}}, \ and\
  \bibinfo {author} {\bibfnamefont {S.}~\bibnamefont {Fan}},\ }\href {\doibase
  10.1038/nphoton.2012.236} {\bibfield  {journal} {\bibinfo  {journal} {Nature
  Photonics}\ }\textbf {\bibinfo {volume} {6}},\ \bibinfo {pages} {782}
  (\bibinfo {year} {2012})}\BibitemShut {NoStop}%
\bibitem [{\citenamefont {Fitzpatrick}\ \emph {et~al.}(2017)\citenamefont
  {Fitzpatrick}, \citenamefont {Sundaresan}, \citenamefont {Li}, \citenamefont
  {Koch},\ and\ \citenamefont {Houck}}]{H1}%
  \BibitemOpen
  \bibfield  {author} {\bibinfo {author} {\bibfnamefont {M.}~\bibnamefont
  {Fitzpatrick}}, \bibinfo {author} {\bibfnamefont {N.~M.}\ \bibnamefont
  {Sundaresan}}, \bibinfo {author} {\bibfnamefont {A.~C.}\ \bibnamefont {Li}},
  \bibinfo {author} {\bibfnamefont {J.}~\bibnamefont {Koch}}, \ and\ \bibinfo
  {author} {\bibfnamefont {A.~A.}\ \bibnamefont {Houck}},\ }\href {\doibase
  10.1103/physrevx.7.011016} {\bibfield  {journal} {\bibinfo  {journal}
  {Physical Review X}\ }\textbf {\bibinfo {volume} {7}},\ \bibinfo {pages}
  {011016} (\bibinfo {year} {2017})}\BibitemShut {NoStop}%
\bibitem [{\citenamefont {Collodo}\ \emph {et~al.}(2019)\citenamefont
  {Collodo}, \citenamefont {Poto{\v{c}}nik}, \citenamefont {Gasparinetti},
  \citenamefont {Besse}, \citenamefont {Pechal}, \citenamefont {Sameti},
  \citenamefont {Hartmann}, \citenamefont {Wallraff},\ and\ \citenamefont
  {Eichler}}]{H2}%
  \BibitemOpen
  \bibfield  {author} {\bibinfo {author} {\bibfnamefont {M.~C.}\ \bibnamefont
  {Collodo}}, \bibinfo {author} {\bibfnamefont {A.}~\bibnamefont
  {Poto{\v{c}}nik}}, \bibinfo {author} {\bibfnamefont {S.}~\bibnamefont
  {Gasparinetti}}, \bibinfo {author} {\bibfnamefont {J.-C.}\ \bibnamefont
  {Besse}}, \bibinfo {author} {\bibfnamefont {M.}~\bibnamefont {Pechal}},
  \bibinfo {author} {\bibfnamefont {M.}~\bibnamefont {Sameti}}, \bibinfo
  {author} {\bibfnamefont {M.~J.}\ \bibnamefont {Hartmann}}, \bibinfo {author}
  {\bibfnamefont {A.}~\bibnamefont {Wallraff}}, \ and\ \bibinfo {author}
  {\bibfnamefont {C.}~\bibnamefont {Eichler}},\ }\href {\doibase
  10.1103/physrevlett.122.183601} {\bibfield  {journal} {\bibinfo  {journal}
  {Physical Review Letters}\ }\textbf {\bibinfo {volume} {122}},\ \bibinfo
  {pages} {183601} (\bibinfo {year} {2019})}\BibitemShut {NoStop}%
\bibitem [{\citenamefont {Wulschner}\ \emph {et~al.}(2016)\citenamefont
  {Wulschner}, \citenamefont {Goetz}, \citenamefont {Koessel}, \citenamefont
  {Hoffmann}, \citenamefont {Baust}, \citenamefont {Eder}, \citenamefont
  {Fischer}, \citenamefont {Haeberlein}, \citenamefont {Schwarz}, \citenamefont
  {Pernpeintner}, \citenamefont {Xie}, \citenamefont {Zhong}, \citenamefont
  {Zollitsch}, \citenamefont {Peropadre}, \citenamefont {Ripoll}, \citenamefont
  {Solano}, \citenamefont {Fedorov}, \citenamefont {Menzel}, \citenamefont
  {Deppe}, \citenamefont {Marx},\ and\ \citenamefont {Gross}}]{H3}%
  \BibitemOpen
  \bibfield  {author} {\bibinfo {author} {\bibfnamefont {F.}~\bibnamefont
  {Wulschner}}, \bibinfo {author} {\bibfnamefont {J.}~\bibnamefont {Goetz}},
  \bibinfo {author} {\bibfnamefont {F.~R.}\ \bibnamefont {Koessel}}, \bibinfo
  {author} {\bibfnamefont {E.}~\bibnamefont {Hoffmann}}, \bibinfo {author}
  {\bibfnamefont {A.}~\bibnamefont {Baust}}, \bibinfo {author} {\bibfnamefont
  {P.}~\bibnamefont {Eder}}, \bibinfo {author} {\bibfnamefont {M.}~\bibnamefont
  {Fischer}}, \bibinfo {author} {\bibfnamefont {M.}~\bibnamefont {Haeberlein}},
  \bibinfo {author} {\bibfnamefont {M.~J.}\ \bibnamefont {Schwarz}}, \bibinfo
  {author} {\bibfnamefont {M.}~\bibnamefont {Pernpeintner}}, \bibinfo {author}
  {\bibfnamefont {E.}~\bibnamefont {Xie}}, \bibinfo {author} {\bibfnamefont
  {L.}~\bibnamefont {Zhong}}, \bibinfo {author} {\bibfnamefont {C.~W.}\
  \bibnamefont {Zollitsch}}, \bibinfo {author} {\bibfnamefont {B.}~\bibnamefont
  {Peropadre}}, \bibinfo {author} {\bibfnamefont {J.-J.~G.}\ \bibnamefont
  {Ripoll}}, \bibinfo {author} {\bibfnamefont {E.}~\bibnamefont {Solano}},
  \bibinfo {author} {\bibfnamefont {K.~G.}\ \bibnamefont {Fedorov}}, \bibinfo
  {author} {\bibfnamefont {E.~P.}\ \bibnamefont {Menzel}}, \bibinfo {author}
  {\bibfnamefont {F.}~\bibnamefont {Deppe}}, \bibinfo {author} {\bibfnamefont
  {A.}~\bibnamefont {Marx}}, \ and\ \bibinfo {author} {\bibfnamefont
  {R.}~\bibnamefont {Gross}},\ }\href {\doibase
  10.1140/epjqt/s40507-016-0048-2} {\bibfield  {journal} {\bibinfo  {journal}
  {{EPJ} Quantum Technology}\ }\textbf {\bibinfo {volume} {3}} (\bibinfo {year}
  {2016}),\ 10.1140/epjqt/s40507-016-0048-2}\BibitemShut {NoStop}%
\bibitem [{\citenamefont {Li}\ \emph {et~al.}(2022)\citenamefont {Li},
  \citenamefont {Yefremenko}, \citenamefont {Lisovenko}, \citenamefont
  {Trevillian}, \citenamefont {Polakovic}, \citenamefont {Cecil}, \citenamefont
  {Barry}, \citenamefont {Pearson}, \citenamefont {Divan}, \citenamefont
  {Tyberkevych}, \citenamefont {Chang}, \citenamefont {Welp}, \citenamefont
  {Kwok},\ and\ \citenamefont {Novosad}}]{C3}%
  \BibitemOpen
  \bibfield  {author} {\bibinfo {author} {\bibfnamefont {Y.}~\bibnamefont
  {Li}}, \bibinfo {author} {\bibfnamefont {V.~G.}\ \bibnamefont {Yefremenko}},
  \bibinfo {author} {\bibfnamefont {M.}~\bibnamefont {Lisovenko}}, \bibinfo
  {author} {\bibfnamefont {C.}~\bibnamefont {Trevillian}}, \bibinfo {author}
  {\bibfnamefont {T.}~\bibnamefont {Polakovic}}, \bibinfo {author}
  {\bibfnamefont {T.~W.}\ \bibnamefont {Cecil}}, \bibinfo {author}
  {\bibfnamefont {P.~S.}\ \bibnamefont {Barry}}, \bibinfo {author}
  {\bibfnamefont {J.}~\bibnamefont {Pearson}}, \bibinfo {author} {\bibfnamefont
  {R.}~\bibnamefont {Divan}}, \bibinfo {author} {\bibfnamefont
  {V.}~\bibnamefont {Tyberkevych}}, \bibinfo {author} {\bibfnamefont {C.~L.}\
  \bibnamefont {Chang}}, \bibinfo {author} {\bibfnamefont {U.}~\bibnamefont
  {Welp}}, \bibinfo {author} {\bibfnamefont {W.-K.}\ \bibnamefont {Kwok}}, \
  and\ \bibinfo {author} {\bibfnamefont {V.}~\bibnamefont {Novosad}},\ }\href
  {\doibase 10.1103/PhysRevLett.128.047701} {\bibfield  {journal} {\bibinfo
  {journal} {Phys. Rev. Lett.}\ }\textbf {\bibinfo {volume} {128}},\ \bibinfo
  {pages} {047701} (\bibinfo {year} {2022})}\BibitemShut {NoStop}%
\bibitem [{\citenamefont {Wang}\ \emph {et~al.}(2021)\citenamefont {Wang},
  \citenamefont {Liu}, \citenamefont {Kockum}, \citenamefont {Li},\ and\
  \citenamefont {Nori}}]{Wangxin}%
  \BibitemOpen
  \bibfield  {author} {\bibinfo {author} {\bibfnamefont {X.}~\bibnamefont
  {Wang}}, \bibinfo {author} {\bibfnamefont {T.}~\bibnamefont {Liu}}, \bibinfo
  {author} {\bibfnamefont {A.~F.}\ \bibnamefont {Kockum}}, \bibinfo {author}
  {\bibfnamefont {H.-R.}\ \bibnamefont {Li}}, \ and\ \bibinfo {author}
  {\bibfnamefont {F.}~\bibnamefont {Nori}},\ }\href {\doibase
  10.1103/PhysRevLett.126.043602} {\bibfield  {journal} {\bibinfo  {journal}
  {Phys. Rev. Lett.}\ }\textbf {\bibinfo {volume} {126}},\ \bibinfo {pages}
  {043602} (\bibinfo {year} {2021})}\BibitemShut {NoStop}%
\bibitem [{\citenamefont {Yu}\ \emph {et~al.}(2021)\citenamefont {Yu},
  \citenamefont {Wang},\ and\ \citenamefont {Wu}}]{Wang2021}%
  \BibitemOpen
  \bibfield  {author} {\bibinfo {author} {\bibfnamefont {H.}~\bibnamefont
  {Yu}}, \bibinfo {author} {\bibfnamefont {Z.}~\bibnamefont {Wang}}, \ and\
  \bibinfo {author} {\bibfnamefont {J.-H.}\ \bibnamefont {Wu}},\ }\href
  {\doibase 10.1103/PhysRevA.104.013720} {\bibfield  {journal} {\bibinfo
  {journal} {Phys. Rev. A}\ }\textbf {\bibinfo {volume} {104}},\ \bibinfo
  {pages} {013720} (\bibinfo {year} {2021})}\BibitemShut {NoStop}%
\bibitem [{\citenamefont {Du}\ \emph {et~al.}(2022)\citenamefont {Du},
  \citenamefont {Zhang}, \citenamefont {Wu}, \citenamefont {Kockum},\ and\
  \citenamefont {Li}}]{Du20222}%
  \BibitemOpen
  \bibfield  {author} {\bibinfo {author} {\bibfnamefont {L.}~\bibnamefont
  {Du}}, \bibinfo {author} {\bibfnamefont {Y.}~\bibnamefont {Zhang}}, \bibinfo
  {author} {\bibfnamefont {J.-H.}\ \bibnamefont {Wu}}, \bibinfo {author}
  {\bibfnamefont {A.~F.}\ \bibnamefont {Kockum}}, \ and\ \bibinfo {author}
  {\bibfnamefont {Y.}~\bibnamefont {Li}},\ }\href {\doibase
  10.1103/PhysRevLett.128.223602} {\bibfield  {journal} {\bibinfo  {journal}
  {Phys. Rev. Lett.}\ }\textbf {\bibinfo {volume} {128}},\ \bibinfo {pages}
  {223602} (\bibinfo {year} {2022})}\BibitemShut {NoStop}%
\bibitem [{\citenamefont {Koch}\ \emph {et~al.}(2007)\citenamefont {Koch},
  \citenamefont {Yu}, \citenamefont {Gambetta}, \citenamefont {Houck},
  \citenamefont {Schuster}, \citenamefont {Majer}, \citenamefont {Blais},
  \citenamefont {Devoret}, \citenamefont {Girvin},\ and\ \citenamefont
  {Schoelkopf}}]{Koch2007}%
  \BibitemOpen
  \bibfield  {author} {\bibinfo {author} {\bibfnamefont {J.}~\bibnamefont
  {Koch}}, \bibinfo {author} {\bibfnamefont {T.~M.}\ \bibnamefont {Yu}},
  \bibinfo {author} {\bibfnamefont {J.}~\bibnamefont {Gambetta}}, \bibinfo
  {author} {\bibfnamefont {A.~A.}\ \bibnamefont {Houck}}, \bibinfo {author}
  {\bibfnamefont {D.~I.}\ \bibnamefont {Schuster}}, \bibinfo {author}
  {\bibfnamefont {J.}~\bibnamefont {Majer}}, \bibinfo {author} {\bibfnamefont
  {A.}~\bibnamefont {Blais}}, \bibinfo {author} {\bibfnamefont {M.~H.}\
  \bibnamefont {Devoret}}, \bibinfo {author} {\bibfnamefont {S.~M.}\
  \bibnamefont {Girvin}}, \ and\ \bibinfo {author} {\bibfnamefont {R.~J.}\
  \bibnamefont {Schoelkopf}},\ }\href {\doibase 10.1103/physreva.76.042319}
  {\bibfield  {journal} {\bibinfo  {journal} {Physical Review A}\ }\textbf
  {\bibinfo {volume} {76}},\ \bibinfo {pages} {042319} (\bibinfo {year}
  {2007})}\BibitemShut {NoStop}%
\bibitem [{\citenamefont {Andersson}\ \emph {et~al.}(2019)\citenamefont
  {Andersson}, \citenamefont {Suri}, \citenamefont {Guo}, \citenamefont
  {Aref},\ and\ \citenamefont {Delsing}}]{Andersson2019}%
  \BibitemOpen
  \bibfield  {author} {\bibinfo {author} {\bibfnamefont {G.}~\bibnamefont
  {Andersson}}, \bibinfo {author} {\bibfnamefont {B.}~\bibnamefont {Suri}},
  \bibinfo {author} {\bibfnamefont {L.}~\bibnamefont {Guo}}, \bibinfo {author}
  {\bibfnamefont {T.}~\bibnamefont {Aref}}, \ and\ \bibinfo {author}
  {\bibfnamefont {P.}~\bibnamefont {Delsing}},\ }\href {\doibase
  10.1038/s41567-019-0605-6} {\bibfield  {journal} {\bibinfo  {journal} {Nature
  Physics}\ }\textbf {\bibinfo {volume} {15}},\ \bibinfo {pages} {1123}
  (\bibinfo {year} {2019})}\BibitemShut {NoStop}%
\bibitem [{\citenamefont {Kannan}\ \emph {et~al.}(2020)\citenamefont {Kannan},
  \citenamefont {Ruckriegel}, \citenamefont {Campbell}, \citenamefont {Kockum},
  \citenamefont {Braumüller}, \citenamefont {Kim}, \citenamefont {Kjaergaard},
  \citenamefont {Krantz}, \citenamefont {Melville}, \citenamefont
  {Niedzielski}, \citenamefont {Vepsäläinen}, \citenamefont {Winik},
  \citenamefont {Yoder}, \citenamefont {Nori}, \citenamefont {Orlando},
  \citenamefont {Gustavsson},\ and\ \citenamefont {Oliver}}]{Kannan2020}%
  \BibitemOpen
  \bibfield  {author} {\bibinfo {author} {\bibfnamefont {B.}~\bibnamefont
  {Kannan}}, \bibinfo {author} {\bibfnamefont {M.~J.}\ \bibnamefont
  {Ruckriegel}}, \bibinfo {author} {\bibfnamefont {D.~L.}\ \bibnamefont
  {Campbell}}, \bibinfo {author} {\bibfnamefont {A.~F.}\ \bibnamefont
  {Kockum}}, \bibinfo {author} {\bibfnamefont {J.}~\bibnamefont {Braumüller}},
  \bibinfo {author} {\bibfnamefont {D.~K.}\ \bibnamefont {Kim}}, \bibinfo
  {author} {\bibfnamefont {M.}~\bibnamefont {Kjaergaard}}, \bibinfo {author}
  {\bibfnamefont {P.}~\bibnamefont {Krantz}}, \bibinfo {author} {\bibfnamefont
  {A.}~\bibnamefont {Melville}}, \bibinfo {author} {\bibfnamefont {B.~M.}\
  \bibnamefont {Niedzielski}}, \bibinfo {author} {\bibfnamefont
  {A.}~\bibnamefont {Vepsäläinen}}, \bibinfo {author} {\bibfnamefont
  {R.}~\bibnamefont {Winik}}, \bibinfo {author} {\bibfnamefont {J.~L.}\
  \bibnamefont {Yoder}}, \bibinfo {author} {\bibfnamefont {F.}~\bibnamefont
  {Nori}}, \bibinfo {author} {\bibfnamefont {T.~P.}\ \bibnamefont {Orlando}},
  \bibinfo {author} {\bibfnamefont {S.}~\bibnamefont {Gustavsson}}, \ and\
  \bibinfo {author} {\bibfnamefont {W.~D.}\ \bibnamefont {Oliver}},\ }\href
  {\doibase 10.1038/s41586-020-2529-9} {\bibfield  {journal} {\bibinfo
  {journal} {Nature}\ }\textbf {\bibinfo {volume} {583}},\ \bibinfo {pages}
  {775} (\bibinfo {year} {2020})}\BibitemShut {NoStop}%
\bibitem [{\citenamefont {Felicetti}\ \emph {et~al.}(2014)\citenamefont
  {Felicetti}, \citenamefont {Sanz}, \citenamefont {Lamata}, \citenamefont
  {Romero}, \citenamefont {Johansson}, \citenamefont {Delsing},\ and\
  \citenamefont {Solano}}]{Felicetti2014}%
  \BibitemOpen
  \bibfield  {author} {\bibinfo {author} {\bibfnamefont {S.}~\bibnamefont
  {Felicetti}}, \bibinfo {author} {\bibfnamefont {M.}~\bibnamefont {Sanz}},
  \bibinfo {author} {\bibfnamefont {L.}~\bibnamefont {Lamata}}, \bibinfo
  {author} {\bibfnamefont {G.}~\bibnamefont {Romero}}, \bibinfo {author}
  {\bibfnamefont {G.}~\bibnamefont {Johansson}}, \bibinfo {author}
  {\bibfnamefont {P.}~\bibnamefont {Delsing}}, \ and\ \bibinfo {author}
  {\bibfnamefont {E.}~\bibnamefont {Solano}},\ }\href {\doibase
  10.1103/physrevlett.113.093602} {\bibfield  {journal} {\bibinfo  {journal}
  {Physical Review Letters}\ }\textbf {\bibinfo {volume} {113}},\ \bibinfo
  {pages} {093602} (\bibinfo {year} {2014})}\BibitemShut {NoStop}%
\bibitem [{\citenamefont {Lu}\ \emph {et~al.}(2017)\citenamefont {Lu},
  \citenamefont {Chakram}, \citenamefont {Leung}, \citenamefont {Earnest},
  \citenamefont {Naik}, \citenamefont {Huang}, \citenamefont {Groszkowski},
  \citenamefont {Kapit}, \citenamefont {Koch},\ and\ \citenamefont
  {Schuster}}]{Lu2017}%
  \BibitemOpen
  \bibfield  {author} {\bibinfo {author} {\bibfnamefont {Y.}~\bibnamefont
  {Lu}}, \bibinfo {author} {\bibfnamefont {S.}~\bibnamefont {Chakram}},
  \bibinfo {author} {\bibfnamefont {N.}~\bibnamefont {Leung}}, \bibinfo
  {author} {\bibfnamefont {N.}~\bibnamefont {Earnest}}, \bibinfo {author}
  {\bibfnamefont {R.}~\bibnamefont {Naik}}, \bibinfo {author} {\bibfnamefont
  {Z.}~\bibnamefont {Huang}}, \bibinfo {author} {\bibfnamefont
  {P.}~\bibnamefont {Groszkowski}}, \bibinfo {author} {\bibfnamefont
  {E.}~\bibnamefont {Kapit}}, \bibinfo {author} {\bibfnamefont
  {J.}~\bibnamefont {Koch}}, \ and\ \bibinfo {author} {\bibfnamefont {D.~I.}\
  \bibnamefont {Schuster}},\ }\href {\doibase 10.1103/physrevlett.119.150502}
  {\bibfield  {journal} {\bibinfo  {journal} {Physical Review Letters}\
  }\textbf {\bibinfo {volume} {119}},\ \bibinfo {pages} {150502} (\bibinfo
  {year} {2017})}\BibitemShut {NoStop}%
\bibitem [{\citenamefont {Leek}\ \emph {et~al.}(2010)\citenamefont {Leek},
  \citenamefont {Baur}, \citenamefont {Fink}, \citenamefont {Bianchetti},
  \citenamefont {Steffen}, \citenamefont {Filipp},\ and\ \citenamefont
  {Wallraff}}]{Leek2010}%
  \BibitemOpen
  \bibfield  {author} {\bibinfo {author} {\bibfnamefont {P.~J.}\ \bibnamefont
  {Leek}}, \bibinfo {author} {\bibfnamefont {M.}~\bibnamefont {Baur}}, \bibinfo
  {author} {\bibfnamefont {J.~M.}\ \bibnamefont {Fink}}, \bibinfo {author}
  {\bibfnamefont {R.}~\bibnamefont {Bianchetti}}, \bibinfo {author}
  {\bibfnamefont {L.}~\bibnamefont {Steffen}}, \bibinfo {author} {\bibfnamefont
  {S.}~\bibnamefont {Filipp}}, \ and\ \bibinfo {author} {\bibfnamefont
  {A.}~\bibnamefont {Wallraff}},\ }\href {\doibase
  10.1103/PhysRevLett.104.100504} {\bibfield  {journal} {\bibinfo  {journal}
  {Phys. Rev. Lett.}\ }\textbf {\bibinfo {volume} {104}},\ \bibinfo {pages}
  {100504} (\bibinfo {year} {2010})}\BibitemShut {NoStop}%
\bibitem [{\citenamefont {Scigliuzzo}\ \emph {et~al.}(2022)\citenamefont
  {Scigliuzzo}, \citenamefont {Calaj\`o}, \citenamefont {Ciccarello},
  \citenamefont {Perez~Lozano}, \citenamefont {Bengtsson}, \citenamefont
  {Scarlino}, \citenamefont {Wallraff}, \citenamefont {Chang}, \citenamefont
  {Delsing},\ and\ \citenamefont {Gasparinetti}}]{PhysRevX.12.031036}%
  \BibitemOpen
  \bibfield  {author} {\bibinfo {author} {\bibfnamefont {M.}~\bibnamefont
  {Scigliuzzo}}, \bibinfo {author} {\bibfnamefont {G.}~\bibnamefont
  {Calaj\`o}}, \bibinfo {author} {\bibfnamefont {F.}~\bibnamefont
  {Ciccarello}}, \bibinfo {author} {\bibfnamefont {D.}~\bibnamefont
  {Perez~Lozano}}, \bibinfo {author} {\bibfnamefont {A.}~\bibnamefont
  {Bengtsson}}, \bibinfo {author} {\bibfnamefont {P.}~\bibnamefont {Scarlino}},
  \bibinfo {author} {\bibfnamefont {A.}~\bibnamefont {Wallraff}}, \bibinfo
  {author} {\bibfnamefont {D.}~\bibnamefont {Chang}}, \bibinfo {author}
  {\bibfnamefont {P.}~\bibnamefont {Delsing}}, \ and\ \bibinfo {author}
  {\bibfnamefont {S.}~\bibnamefont {Gasparinetti}},\ }\href {\doibase
  10.1103/PhysRevX.12.031036} {\bibfield  {journal} {\bibinfo  {journal} {Phys.
  Rev. X}\ }\textbf {\bibinfo {volume} {12}},\ \bibinfo {pages} {031036}
  (\bibinfo {year} {2022})}\BibitemShut {NoStop}%
\bibitem [{\citenamefont {Walgate}\ \emph {et~al.}(2000)\citenamefont
  {Walgate}, \citenamefont {Short}, \citenamefont {Hardy},\ and\ \citenamefont
  {Vedral}}]{Walgate2000}%
  \BibitemOpen
  \bibfield  {author} {\bibinfo {author} {\bibfnamefont {J.}~\bibnamefont
  {Walgate}}, \bibinfo {author} {\bibfnamefont {A.~J.}\ \bibnamefont {Short}},
  \bibinfo {author} {\bibfnamefont {L.}~\bibnamefont {Hardy}}, \ and\ \bibinfo
  {author} {\bibfnamefont {V.}~\bibnamefont {Vedral}},\ }\href {\doibase
  10.1103/PhysRevLett.85.4972} {\bibfield  {journal} {\bibinfo  {journal}
  {Phys. Rev. Lett.}\ }\textbf {\bibinfo {volume} {85}},\ \bibinfo {pages}
  {4972} (\bibinfo {year} {2000})}\BibitemShut {NoStop}%
\bibitem [{\citenamefont {Maniscalco}\ \emph {et~al.}(2008)\citenamefont
  {Maniscalco}, \citenamefont {Francica}, \citenamefont {Zaffino},
  \citenamefont {Gullo},\ and\ \citenamefont {Plastina}}]{Maniscalco2008}%
  \BibitemOpen
  \bibfield  {author} {\bibinfo {author} {\bibfnamefont {S.}~\bibnamefont
  {Maniscalco}}, \bibinfo {author} {\bibfnamefont {F.}~\bibnamefont
  {Francica}}, \bibinfo {author} {\bibfnamefont {R.~L.}\ \bibnamefont
  {Zaffino}}, \bibinfo {author} {\bibfnamefont {N.~L.}\ \bibnamefont {Gullo}},
  \ and\ \bibinfo {author} {\bibfnamefont {F.}~\bibnamefont {Plastina}},\
  }\href {\doibase 10.1103/physrevlett.100.090503} {\bibfield  {journal}
  {\bibinfo  {journal} {Physical Review Letters}\ }\textbf {\bibinfo {volume}
  {100}},\ \bibinfo {pages} {090503} (\bibinfo {year} {2008})}\BibitemShut
  {NoStop}%
\bibitem [{\citenamefont {Birnbaum}\ \emph {et~al.}(2005)\citenamefont
  {Birnbaum}, \citenamefont {Boca}, \citenamefont {Miller}, \citenamefont
  {Boozer}, \citenamefont {Northup},\ and\ \citenamefont {Kimble}}]{H8}%
  \BibitemOpen
  \bibfield  {author} {\bibinfo {author} {\bibfnamefont {K.~M.}\ \bibnamefont
  {Birnbaum}}, \bibinfo {author} {\bibfnamefont {A.}~\bibnamefont {Boca}},
  \bibinfo {author} {\bibfnamefont {R.}~\bibnamefont {Miller}}, \bibinfo
  {author} {\bibfnamefont {A.~D.}\ \bibnamefont {Boozer}}, \bibinfo {author}
  {\bibfnamefont {T.~E.}\ \bibnamefont {Northup}}, \ and\ \bibinfo {author}
  {\bibfnamefont {H.~J.}\ \bibnamefont {Kimble}},\ }\href {\doibase
  10.1038/nature03804} {\bibfield  {journal} {\bibinfo  {journal} {Nature}\
  }\textbf {\bibinfo {volume} {436}},\ \bibinfo {pages} {87} (\bibinfo {year}
  {2005})}\BibitemShut {NoStop}%
\bibitem [{\citenamefont {Faraon}\ \emph {et~al.}(2008)\citenamefont {Faraon},
  \citenamefont {Fushman}, \citenamefont {Englund}, \citenamefont {Stoltz},
  \citenamefont {Petroff},\ and\ \citenamefont {Vu{\v{c}}kovi{\'{c}}}}]{H9}%
  \BibitemOpen
  \bibfield  {author} {\bibinfo {author} {\bibfnamefont {A.}~\bibnamefont
  {Faraon}}, \bibinfo {author} {\bibfnamefont {I.}~\bibnamefont {Fushman}},
  \bibinfo {author} {\bibfnamefont {D.}~\bibnamefont {Englund}}, \bibinfo
  {author} {\bibfnamefont {N.}~\bibnamefont {Stoltz}}, \bibinfo {author}
  {\bibfnamefont {P.}~\bibnamefont {Petroff}}, \ and\ \bibinfo {author}
  {\bibfnamefont {J.}~\bibnamefont {Vu{\v{c}}kovi{\'{c}}}},\ }\href {\doibase
  10.1038/nphys1078} {\bibfield  {journal} {\bibinfo  {journal} {Nature
  Physics}\ }\textbf {\bibinfo {volume} {4}},\ \bibinfo {pages} {859} (\bibinfo
  {year} {2008})}\BibitemShut {NoStop}%
\end{thebibliography}%

\end{document}